\documentclass[prx,twocolumn,showpacs,amsmath,amssymb,superscriptaddress]{revtex4-1}
\usepackage{epsfig}
\usepackage{graphicx}
\usepackage{hyperref}
\usepackage{color}
\usepackage{tabu}
\makeatother
\usepackage{pifont}
\usepackage{multirow}
\usepackage[table,xcdraw]{xcolor}
\usepackage{booktabs}
\usepackage{bm}

\begin{document}

\title{Emerging research landscape of altermagnetism}
\author{Libor \v{S}mejkal}
\affiliation{Institut f\"ur Physik, Johannes Gutenberg Universit\"at Mainz, D-55099 Mainz, Germany}
\affiliation{Institute of Physics, Czech Academy of Sciences, Cukrovarnick\'{a} 10, 162 00 Praha 6 Czech Republic}

\author{Jairo Sinova}
\affiliation{Institut f\"ur Physik, Johannes Gutenberg Universit\"at Mainz, D-55099 Mainz, Germany}
\affiliation{Institute of Physics, Czech Academy of Sciences, Cukrovarnick\'{a} 10, 162 00 Praha 6 Czech Republic}

\author{Tomas Jungwirth}
\affiliation{Institute of Physics, Czech Academy of Sciences, Cukrovarnick\'{a} 10, 162 00 Praha 6 Czech Republic}
\affiliation{School of Physics and Astronomy, University of Nottingham, Nottingham NG7 2RD, United Kingdom}

\date{\today}

\begin{abstract}
Magnetism is one of the largest, most fundamental, and technologically most relevant fields of condensed-matter physics. Traditionally, two basic magnetic phases have been considered -- ferromagnetism and antiferromagnetism. The breaking of the time-reversal symmetry and spin splitting of the electronic states by the magnetization in ferromagnets underpins a range of macroscopic responses in this extensively explored and exploited type of magnets. By comparison, antiferromagnets have vanishing net magnetization.  This Perspective reflects on recent observations of materials hosting an intriguing ferromagnetic-antiferromagnetic dichotomy, in which spin-split spectra and macroscopic observables, akin to ferromagnets, are accompanied by antiparallel magnetic order with vanishing magnetization, typical of antiferromagnets. An unconventional non-relativistic symmetry-group formalism offers a resolution of this apparent contradiction by delimiting a third basic magnetic phase, dubbed altermagnetism. Our Perspective starts with an overview of the still emerging  unique phenomenology of the phase, and of the wide array of altermagnetic material candidates. In the main part of the article, we illustrate how altermagnetism can enrich our understanding of overarching condensed-matter physics concepts, and have impact on prominent condensed-matter research areas. 
\end{abstract}
\maketitle

\tableofcontents

\twocolumngrid

\section{Introduction}
Magnetic solids are traditionally divided into two elementary phases -- ferromagnets and antiferromagnets  \cite{Neel1971}. Ferromagnets, known for several millennia, are characterized by a strong macroscopic magnetization. They generate a range of macroscopic phenomena originating from the spin-split electronic band structure with broken time-reversal ($\cal{T}$) symmetry, induced by the net magnetization. Antiferromagnets, on the other hand, were discovered only a century ago, due  to their vanishing net magnetization that makes them behave in many aspects as non-magnetic materials. In a traditional picture of antiferromagnetism,  a compensating antiparallel ordering of atomic magnetic moments, i.e. the effective cancellation of atomic moments leading to the vanishingly small macroscopic net magnetization, has been thought to generate no spin splitting of electronic states, and to be invisible to the macroscopic electrical or optical probes commonly used in ferromagnets. 

\begin{figure}[h!]
\begin{center}
\hspace*{-0cm}\epsfig{width=.78\columnwidth,angle=0,file=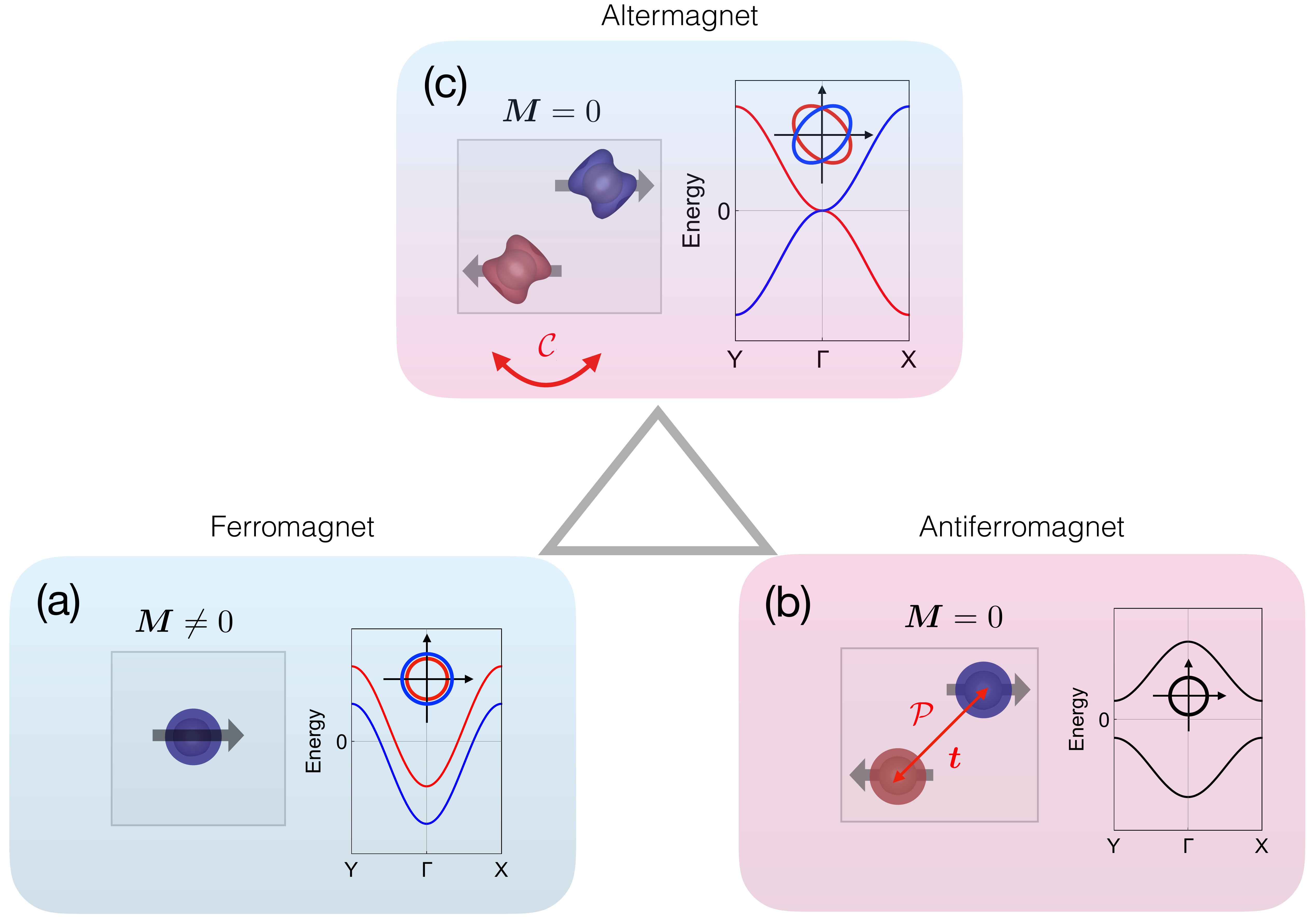}
\end{center}
\vspace{-.47cm}
\caption{Illustrative model distinction between conventional collinear  ferromagnetic and antiferromagnetic phases, and the emerging altermagnetic phase, highlighted  in the crystal-structure real space and  electronic-structure momentum space. (a) Ferromagnetic model with one spin sublattice and corresponding magnetization (left panel), momentum-independent spin splitting (right panel) and isotropic spin-split Fermi surfaces (inset of right panel). (b) Antiferromagnetic model with opposite-spin sublattices connected by translation ($\bm t$) or inversion ($\cal{P}$) transformations, and corresponding zero net magnetization and spin-degenerate bands. (c) Altermagnetic model with opposite-spin sublattices connected by rotation ($C$) transformation, and corresponding zero net magnetization, anisotropic sublattice spin densities, spin splitting with alternating sign, and anisotropic spin-split Fermi surfaces.
}
\label{FM_AF_AM}
\end{figure}

Recently, diverse condensed matter research communities have been intrigued by theoretical predictions of  $\cal{T}$-symmetry breaking macroscopic phenomena \cite{Smejkal2021b,Smejkal2020,Naka2019,Feng2020a,Reichlova2020,Mazin2021,Gonzalez-Hernandez2021,Naka2021,Bose2021,Bai2021,Karube2021,Smejkal2022,Shao2021,Ma2021,Smejkal2021a} and spin-split band structures \cite{Smejkal2021b,Noda2016,Smejkal2020,Ahn2019,Hayami2019,Naka2019,Yuan2020,Feng2020a,Hayami2020,Reichlova2020,Yuan2021,Egorov2021,Mazin2021,Gonzalez-Hernandez2021,Naka2021,Smejkal2022,Shao2021,Ma2021,Smejkal2021a,Liu2021,Yang2021}, that are typical of ferromagnets, in materials with compensated antiparallel magnetic ordering, that is characteristic of antiferromagnets.  The apparent ferromagnetic-antiferromagnetic  dichotomy in these materials challenges the traditional division of materials by the two basic magnetic phases. A recent development \cite{Smejkal2021a} exploiting an unconventional non-relativistic symmetry-group  formalism \cite{Brinkman1966,Litvin1974,Litvin1977} can resolve this contradiction by delimiting, apart from the traditional ferromagnetism and antiferromagnetism, a third distinct and comparably abundant phase.  This third phase, dubbed altermagnetism, is  characterized by a compensated magnetic order with opposite-spin sublattices connected by crystal-rotation symmetries, and by  band structures with broken $\cal{T}$-symmetry and alternating sign of the spin splitting in the momentum space. The distinction of the three phases is highlighted in Fig.~\ref{FM_AF_AM}.

\onecolumngrid


\twocolumngrid

As we will discuss in this Perspective, altermagnetism emerges on the basic level of the crystal-potential theory and effective single-particle non-relativistic description of band structures of collinear magnets. It is, therefore, a robust elementary magnetic phase. The theoretical prediction of altermagnetism thus complements in a unique fundamental way modern studies of spin quantum phases associated with more complex, and often more subtle, topological phenomena, many-body correlations, relativistic physics, or frustrated magnetic interactions \cite{Haldane2017,Bednorz1988,Stormer1999a,Nagaosa2010,Sinova2015,Jungwirth2016,Manchon2019,Franz2013,Tokura2019,Nakatsuji2022}.

Altermagnetism is expected to be abundant in nature and to occur in both three-dimensional and two-dimensional crystals, in diverse structural or chemistry types, and in conduction types covering the whole spectrum from insulators to superconductors. In Sec.~II, we give an overview of the predicted characteristic features, symmetries and material landscape of the altermagnetic phase.

The properties connected to the spin-split $\cal{T}$-symmetry broken band structures of altermagnets open up a potential for previously unforeseeable developments in a broad condensed-matter physics field. In Sec.~III, we  highlight the distinct properties of altermagnets in the context of overarching physical concepts of Kramers theorem, Fermi-liquid instabilities, electron and magnon quasiparticles, and Berry phase and non-dissipative transport. Sec.~IV then outlines  foreseen potential of altermagnets in selected  active research areas, including spintronics, ultra-fast optics, neuromorphics, thermoelectrics, field-effect electronics, multiferroics, and high-temperature superconductivity.

While our focus in the following sections is on the emerging field  of altermagnetism from the theory perspective, we point out here that first measurements have already indicated that altermagnetism can soon become an active experimental field. Shortly after the theory predictions of  the possible coexistence of the compensated antiparallel magnetic order and the $\cal{T}$-symmetry breaking macroscopic phenomena, a supporting evidence  has been brought up  by initial experiments \cite{Feng2020a,Bose2021,Bai2021,Karube2021}, as highlighted in Tab.~I.  Apart from the fundamental physics interests, we expect that intense experimental research  will be also driven by the potential impact of altermagnetism on technology. Altermagnetism can occur in crystals with common light elements, high magnetic ordering temperatures and strong spin-coherence, that are among the key prerequisites for practical device applications.

\newpage

\onecolumngrid

\begin{table}[h!]
\begin{tabular}{ccccc}
\textbf{Macroscopic response\;\;\;}                                     & \textbf{Theory in RuO$_2$\;\;\;}                                               & \textbf{Experiment in RuO$_2$} & \textbf{Theory in other materials}                                                                                                                                                                                                                                                                                           \\ \hline \hline \\
\textbf{Anomalous Hall effect}    &  2019 \cite{Smejkal2020} & 2020 \cite{Feng2020a} & SrRuO$_3$ \cite{Samanta2020},                                                                                                                                                                                                                                                            Mn$_5$Si$_3$ \cite{Reichlova2020}, 
$\kappa$-Cl \cite{Naka2020}, (Cr,Fe)Sb$_2$ \cite{Mazin2021}       \\                                                                                                                                                                                                                                                                                                                  \\ \hline \\
\textbf{Spin current and torque}                             & 2020 \cite{Gonzalez-Hernandez2021,Smejkal2021a} & 2021 \cite{Bose2021,Bai2021,Karube2021}          & $\kappa$-Cl \cite{Naka2019}, CaCrO$_3$ \cite{Naka2020a} \\  \\ \hline
\end{tabular}
\caption{Theoretical predictions, supported by experiments, of $\cal{T}$-symmetry breaking macroscopic phenomena in  RuO$_2$, and a list of other altermagnetic materials in which these macroscopic responses were also theoretically predicted. The anomalous Hall effect is a $\cal{T}$-odd off-diagonal component of the electrical conductivity tensor \cite{Smejkal2021b}. 
The $\cal{T}$-odd spin current can be generated along or transverse to an applied electrical bias; an out-of-plane spin current generated  by an in-plane electrical bias in an altermagnetic layer can exert a torque on magnetic moments in an adjacent layer in a multilayer stack \cite{Gonzalez-Hernandez2021}.}
\label{exp}
\end{table}

\begin{figure}[h!]
\begin{center}
\hspace*{-0cm}\epsfig{width=.9\columnwidth,angle=0,file=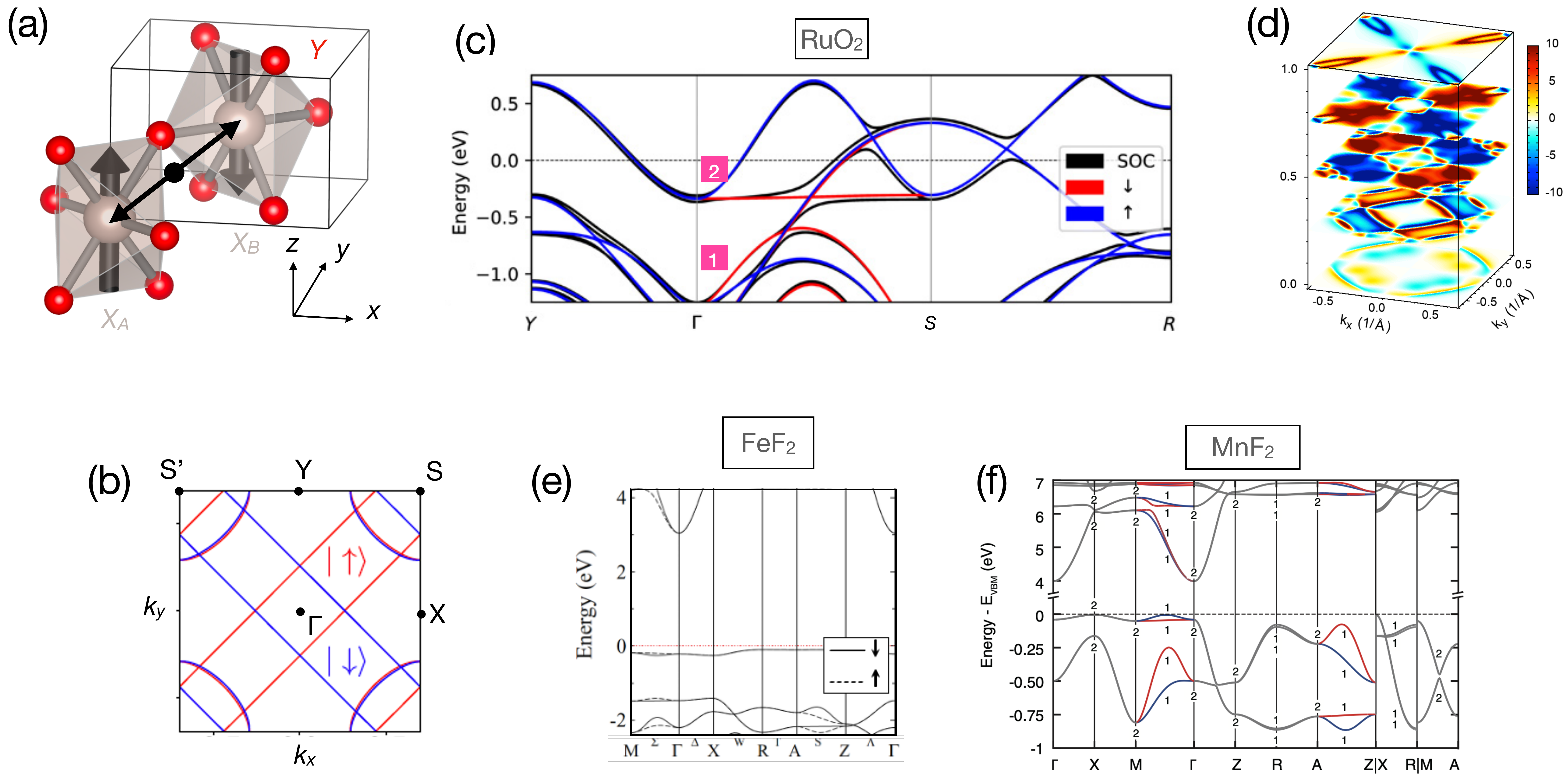}
\end{center}
\vspace{-.5cm}
\caption{
(a) Schematics of the rutile XY$_2$ crystal structure with antiparallel magnetic moments on ${\rm X_A}$ and ${\rm X_B}$ magnetic sublattices. (b) Brillouin zone of the rutile crystal and  {\em ab initio} non-relativistic calculation of a wave-vector $k_z = 0$ cut of the anisotropic spin-split Fermi surface of metallic RuO$_2$. (c) {\em Ab initio} altermagnetic spin splitting of bands in RuO$_2$, calculated without (red and blue)  and with (black) relativistic spin-orbit coupling. (d) {\em Ab initio} altermagnetic spin-split Fermi surface for selected $k_z$ values in RuO$_2$ with correlations accounted for within the dynamical mean-field theory. (d,e) {\em Ab initio} altermagnetic spin-split bands of insulating FeF$_2$ and MnF$_2$, respectively. Adapted from Refs.~\cite{Smejkal2020,Ahn2019,Lopez-Moreno2012,Yuan2020}.} 
\label{ab_initio_AM}
\end{figure}

\twocolumngrid

\section{Altermagnetic phase}

\subsection{Ab initio band-structures}

In Fig.~\ref{ab_initio_AM} we show representative non-relativistic band structures of metallic RuO$_2$ \cite{Smejkal2020,Ahn2019,Gonzalez-Hernandez2021,Smejkal2022,Shao2021,Smejkal2021a} and insulating FeF$_2$  \cite{Lopez-Moreno2012}  and MnF$_2$ \cite{Lopez-Moreno2016,Yuan2020}, on which we illustrate key characteristics of the spin-split $\cal{T}$-symmetry broken band structures of altermagnets. These altermagnetic material examples belong to the family of crystals with the rutile structure (Fig.~\ref{ab_initio_AM}a). For several insulating members of the rutile family,  the compensated antiparallel arrangement of magnetic moments (Fig.~\ref{ab_initio_AM}a)  was well known already to N\'eel and his contemporaries who, ironically,  introduced them into the literature as a classic representation of antiferromagnetism \cite{Neel1953,Kurti1988}.  The notion was based on focusing on the lattice of magnetic atoms alone, while omitting the essential role of non-magnetic atoms on magnetism in the rutile crystals. This may be one of the reasons why the spin splitting and $\cal{T}$-symmetry breaking of their non-relativistic band structures (Figs.~\ref{ab_initio_AM}b-f) remained unnoticed for nearly a century. 

Remarkably, the room-temperature antiparallel magnetic ordering in metallic rutile RuO$_2$ was discovered \cite{Berlijn2017a,Zhu2018} and investigated \cite{Lovesey2021,Occhialini2021} only recently. The subsequent theoretical and experimental exploration of the $\cal{T}$-symmetry breaking macroscopic responses \cite{Smejkal2020,Ahn2019,Hayami2019,Feng2020a,Gonzalez-Hernandez2021,Bose2021,Bai2021,Karube2021,Smejkal2022,Shao2021,Smejkal2021a} has made 
RuO$_2$  one of the workhorse materials of the emerging research of altermagnetism.  

Figures~\ref{ab_initio_AM}b-f show that the altermagnetic spin splitting is strongly momentum-dependent in all three rutiles. In RuO$_2$, it  reaches in parts of the Brillouin zone a $\sim$1~eV scale, which is comparable  to the spin-splitting magnitudes in ferromagnets. Unlike ferromagnets, however, the altermagnetic spin splitting in the non-relativistic bands is accompanied by a zero net magnetization. 

Figure~\ref{ab_initio_AM} also illustrates that spin splittings in altermagnets can exceed by an order  of magnitude the record relativistic spin splittings in bulk crystals with heavy elements \cite{Ishizaka2011}. Moreover, unlike the momentum-dependent spin textures in the relativistic bands, spin is a good quantum number and the electronic states share a common momentum-independent spin quantization axis in the non-relativistic bands of altermagnets. 

The spin-split parts of the altermagnetic band structure are complemented by spin degeneracies along certain lines or surfaces in the Brillouin zone. In Sec.~II.B we will show that non-relativistic symmetries of the altermagnetic spin group corresponding to the given crystal allow for 
the spin splittings and  protect the spin degeneracies in high symmetry planes and lines \cite{Smejkal2021a}.  

Fermi surface cuts shown in Fig.~\ref{ab_initio_AM}b,d highlight the typical anisotropic nature of the spin-split Fermi surfaces, with an equal number of states in the opposite spin channels, and with the spin-momentum locking which is even under the inversion of the momentum and breaks $\cal{T}$-symmetry. 

{\em Ab initio} calculations in RuO$_2$ shown in Fig.~\ref{ab_initio_AM}c,d also demonstrate that the altermangetic spin splitting is only weakly affected by the relativistic spin-orbit coupling, and that the prominent features of the altermagnetic spin-momentum locking are preserved when including correlation effects beyond the local-spin-density approximation of the density-functional theory  \cite{Smejkal2020,Ahn2019,Yuan2020}. A stable itinerant altermagnetism is further confirmed in calculations without Hubbard correlations in other materials, such as Mn$_5$Si$_3$ \cite{Reichlova2020} or KRu$_4$O$_8$ \cite{Smejkal2021a}. A sizable altermangetic spin splitting survives also in the presence of a strong alloying disorder, as shown in altermagnetic Cr$_{0.15}$Fe$_{0.85}$Sb$_{2}$ \cite{Mazin2021}. All this can be understood by the fact that the altermagnetic symmetries, discussed in the following section, can in principle hold equally well for the effective single-particle Kohn-Sham potential, as well as for the Dyson-equation description of correlated or disordered systems. It highlights that the altermagnetic phase is robust in a broad range of materials, can be described within the effective single-particle Kohn-Sham theory, and that the non-relativistic 
crystal potential can play a dominant role in both uncorrelated and correlated altermagnets.

\subsection{Symmetry classification and description}
We now move  from the microscopic {\em ab initio} theory to a discussion of altermagnetism from the symmetry perspective. We start with an example comparing the description of RuO$_2$ by relativistic magnetic-group symmetries and by non-relativistic spin-group symmetries \cite{Smejkal2021a}. This will illustrate why we choose the non-relativistic spin-group formalism for the general symmetry classification and description of the altermagnetic phase.

The relativistic magnetic groups  \cite{Landau1984,Shubnikov1964,Tavger1956,Bradley,Litvin2013,Gallego2016a} consider transformations in coupled real physical space and the space of magnetic moment vectors. They have represented a primary symmetry tool for describing magnetic structures in materials' databases \cite{Litvin2013,Gallego2016a}, and have been broadly applied in the research of equilibrium and non-equilibrium magnetic phenomena, including their modern topological variants \cite{Smejkal2018,Watanabe2018c,Xu2020,Elcoro2021,Cano2019,Lenggenhager2022}. 

\begin{figure}[h!]
\begin{center}
\hspace*{-.1cm}\epsfig{width=1.05\columnwidth,angle=0,file=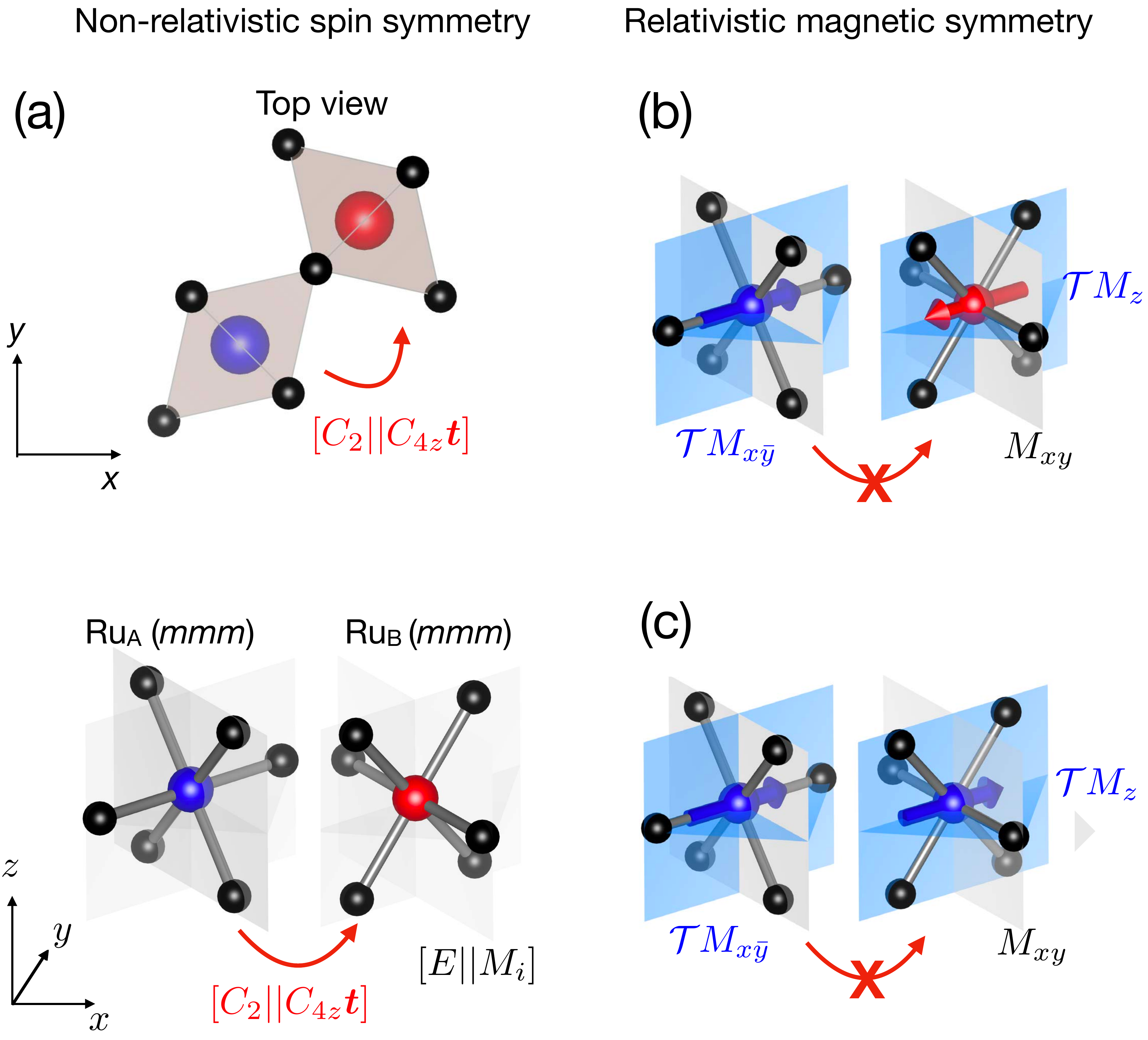}
\end{center}
\vspace{0cm}
\caption{(a) Schematic top view and 3D view of  the 
RuO$_2$ crystal with opposite spin directions on Ru$_{\rm A}$ and Ru$_{\rm B}$ sublattices depicted by red and blue color, oxygen atoms shown in black, and with the depicted non-relativistic spin group symmetries in the notation of Ref.~ \cite{Smejkal2021a} (see text). Curved red arrow and its label highlights the generator of opposite-spin sublattice transformations, and the generators of the halving subgroup of same-spin sublattice transformations are also highlighted (in black). (b,c) Schematic spin arrangement on the 
RuO$_2$  crystal with antiparallel (b) and parallel (c) spin directions and the crystallographic spin-axis orientation depicted by red and blue arrows, and with the depicted relativistic magnetic symmetry group generators. The crossed arrow highlights that the magnetic group contains no opposite spin-sublattice transformation elements. 
(c) The spin-polarized relativistic Fermi surface highlighting the presence of the approximate spin symmetry $\left[ C_{2} \vert \vert C_{4z}\right]$, omitted by the magnetic group. (d) Relativistic momentum-resolved Berry curvature hotspots originating from the avoided crossings along the $k_{x,y}=0$ lines, whose position in the momentum space is determined by the spin symmetries in the absence of the relativistic spin-orbit coupling. 
(e) Fermi surface resolved Berry curvature of FeSb$_2$ altermagnet illustrates pronounced contributions to Berry curvature from the quasi-nodal surface cross $k_{x}=0,k_{y}=0$.  Adapted from Ref.~ \cite{Smejkal2021a}. 
} 
\label{RuO2_symmetries}
\end{figure}

The non-relativistic spin group formalism \cite{Brinkman1966,Litvin1974,Litvin1977} is a generalization of the relativistic magnetic groups because it allows for different transformations to act simultaneously in the decoupled spin and real space. The spin groups have been employed only sporadically in magnetic literature \cite{Andreev1980,Smejkal2021a,Corticelli2022,Liu2021,Yang2021,Turek2022}. 

The non-relativistic spin point group of RuO$_2$ is a direct product, ${\bf r}\times {\bf R}_s^{\rm RuO_2}$, of a spin-only group ${\bf r}=[{\bf C}_\infty\parallel E] + [\bar{C}_2{\bf C}_\infty\parallel \cal{T}]$ and of a non-trivial spin point group ${\bf R}_s^{\rm RuO_2}=[E\parallel{\bf H}]+[{C}_2\parallel C_{4z}] \, [E\parallel{\bf H}]=2_{4/}1_{m}2_{m}1_{m}$ \cite{Smejkal2021a}. In the spin-only group,  ${\bf C}_\infty$  is a group of arbitrary rotations of the spin space around the common axis of spins of the collinear magnet, $E$ is real-space identity,  and $\bar{C}_2$ is  a 180$^\circ$ spin-space rotation transformation (${C}_2$) around an axis perpendicular to the spins  combined with the spin-space inversion (i.e. time-reversal) \cite{Litvin1974,Litvin1977}. 

${\bf H}=mmm$ is a halving subgroup of real-space transformations of the non-magnetic crystallographic group $4/mmm$ of RuO$_2$ \cite{Smejkal2021a}. The subgroup $[E\parallel{\bf H}]$, where $E$ is spin-space identity, contains symmetry transformations which interchange atoms belonging only to one of the two spin sublattices. Its generators, $[E\parallel \bar{E}]$, $[E\parallel M_{xy}]$,  and $[E\parallel M_z]$, are shown in Fig.~\ref{RuO2_symmetries}a. Here $\bar{E}$ is the spin-group notation \cite{Litvin1974,Litvin1977,Smejkal2021a} for real-space inversion ($\cal{P}$), and $M_{xy}$ and  $M_z$ are  real-space mirror transformations.

$[{C}_2\parallel C_{4z}]$, where $C_{4z}$ is a four-fold real-space rotation, is a generator of symmetry transformations which interchange atoms between opposite-spin sublattices, as also shown in Fig.~\ref{RuO2_symmetries}a. The presence of symmetry transformations connecting the opposite-spin sublattices protects the zero net magnetization of the non-relativistic magnetic structure of RuO$_2$. They also determine the separation of opposite-spin equal-energy electronic states in the momentum space, i.e., the symmetry-protected spin-degenerate parts of the Brillouin zone, and the parts where spin splitting is allowed by symmetry. 

The relativistic magnetic point group of RuO$_2$ is 
$m'm'm$
for the spin-axis direction shown in Fig.~\ref{RuO2_symmetries}b. Its generators are  space-inversion, mirror  $M_z$ 
combined with $\cal{T}$-transformation, and mirror $M_{xy}$. The same magnetic group describes a fully compensated antiparallel magnetic order (Fig.~\ref{RuO2_symmetries}b), a parallel magnetic order with a strong non-relativistic ferromagnetic moment (Fig.~\ref{RuO2_symmetries}c), as well as an antiparallel magnetic order with a weak uncompensated relativistic magnetization \cite{Smejkal2021a}. This example illustrates that the relativistic magnetic groups do not generally separate between relativistic and non-relativistic, compensated and non-compensated, or collinear and non-collinear magnetic phases. Magnetic space groups of type II, describing $\cal{T}$-invariant crystals without a magnetic order, are an exception for which a transition to a non-relativistic physics description in decoupled spin and real space can be generally performed by making a direct product with the spin-only group $SU$(2) of arbitrary spin-space rotations \cite{Litvin1974,Suzuki2017}. For the remaining magnetic space groups of type I, III, and IV, generally encompassing crystals with collinear and non-collinear magnetic order \cite{Yuan2020,Yuan2021}, a transition to the non-relativistic physics description is not available \cite{Suzuki2017,Hayami2019,Hayami2020,Smejkal2021a}.

\begin{figure}[h!]
\begin{center}
\hspace*{-0cm}\epsfig{width=1\columnwidth,angle=0,file=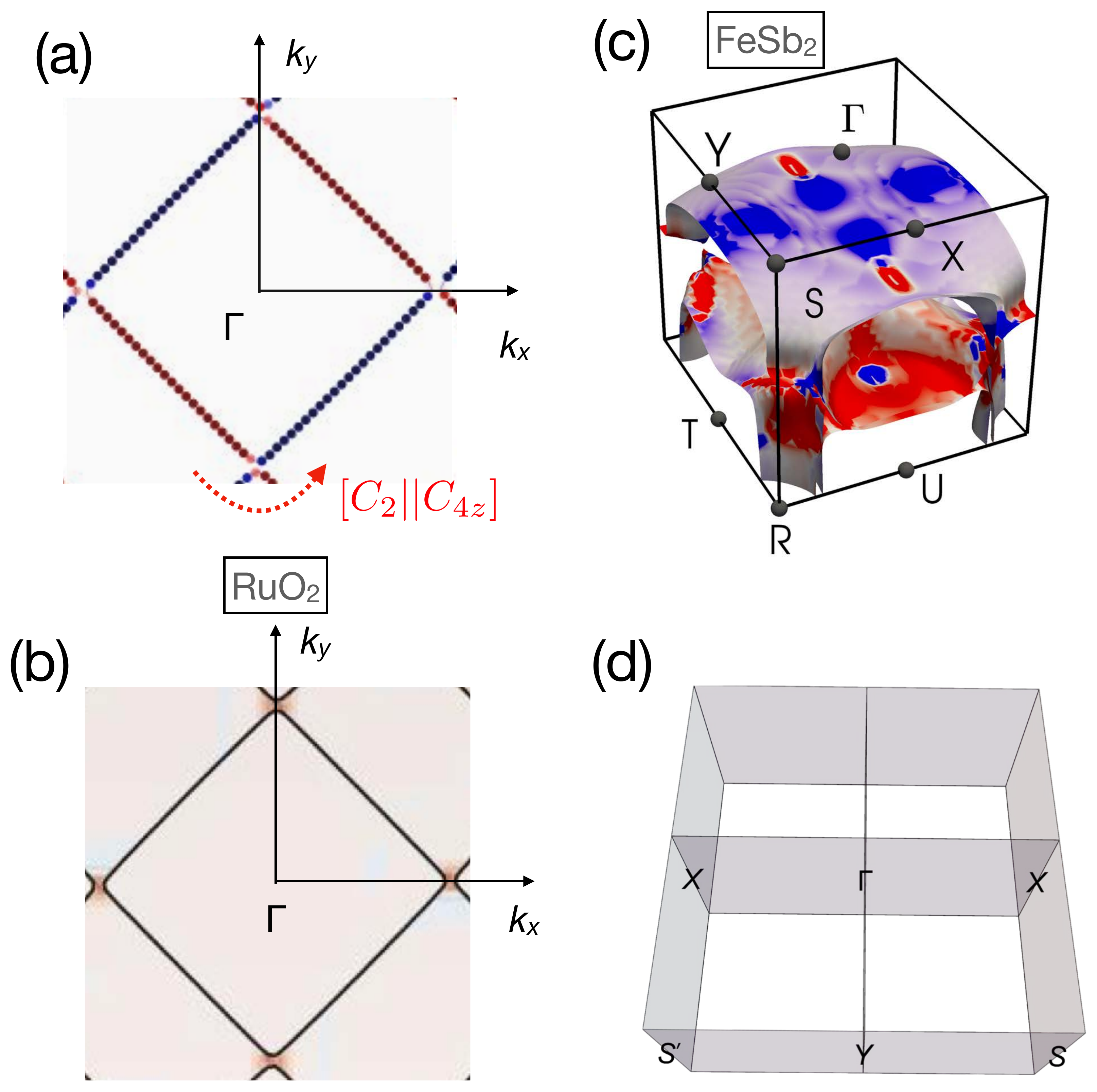}
\end{center}
\vspace{0cm}
\caption{(a) Spin-polarized relativistic Fermi surface highlighting the presence of the approximate spin symmetry $\left[ C_{2} \vert \vert C_{4z}\right]$, omitted by the magnetic group. (b) Relativistic momentum-resolved Berry curvature hotspots originating from the avoided crossings along the $k_{x,y}=0$ lines, whose position in the momentum space is determined by the spin symmetries in the absence of the relativistic spin-orbit coupling. 
(c) Fermi surface resolved Berry curvature of FeSb$_2$ altermagnet illustrates pronounced contributions to Berry curvature from the quasi-nodal surface cross $k_{x}=0,k_{y}=0$.  (d) Brilouin zone notation. Adapted from Ref.~ \cite{Feng2020a,Smejkal2021a,Mazin2021}. 
} 
\label{Berry_curv}
\end{figure}

Figure~\ref{Berry_curv}a shows that the prominent $[{C}_2\parallel C_{4z}]$ symmetry of the non-relativistic spin point group of RuO$_2$ is an apparent  dominant features of the Fermi surfaces even when the relativistic spin-orbit coupling is included in the band-structure calculation. In contrast, the prominent four-fold symmetry is absent in the relativistic magnetic group.  This illustrates that the non-relativistic spin-groups represent an example of approximate, or so called "hidden" symmetries. They allow for a systematic symmetry classification and description of magnetic phases  arising from the typically dominant non-relativistic  crystal potentials \cite{Brinkman1966,Andreev1980,Smejkal2021a,Corticelli2022}.

The spin-group formalism catagorizes all non-relativistic collinear magnetic structures into three distinct phases \cite{Smejkal2021a}: (i) Ferromagnets (ferrimagnets) with one spin lattice (or opposite-spin sublattices not connected by any symmetry transformation), (ii) spin-degenerate antiferromagnets with opposite-spin sublattices connected by translation 
(spin-group symmetry $[{C}_2\parallel {\bm t}]$) 
or inversion 
(spin-group symmetry [${C}_2\parallel \bar{E}]$), and (iii) altermagnets with opposite-spin sublattices connected by  rotation transformations, but not connected by translation or inversion. 

The altermagnetic spin groups have a   general form of the direct product, ${\bf r}\times {\bf R}_s$. Here the spin-only group,
\begin{equation}
{\bf r}=[{\bf C}_\infty\parallel E] + [\bar{C}_2{\bf C}_\infty\parallel \cal{T}],
\label{spin-only-group}
\end{equation}
is common to all three non-relativistic collinear phases, while  the non-trivial spin groups given by,  
\begin{equation}
{\bf R}_s=[E\parallel{\bf H}]+[{C}_2\parallel A] \, [E\parallel{\bf H}],
\label{AMgroup}
\end{equation}
correspond exclusively to  altermagnets. Here $A$ is the real-space rotation transformation (proper or improper, symmorphic or non-symmorphic). Altermagnets have split, but equally populated spin-up and spin-down energy iso-surfaces in the non-relativistic band structure that breaks $\cal{T}$-symmetry. 
This makes altermagnetism distinct from the ferromagnetic phase with a non-zero net magnetization, the spin-degenerate antiferromagnetic phase, or non-magnetic relativistic systems with $\cal{T}$-invariant bands. In Tab.~\ref{SLG} we summarize the properties  derived from the spin group symmetries, which will guide our discussion in the following sections. The first two lines regard symmetries of the spin-only group which apply to all three non-relativistic collinear phases. The remaining lines in Tab.~\ref{SLG} apply only to altermagnets.

\onecolumngrid

\bigskip

\begin{table}[h!]

\begin{tabular}{ccc}
\textbf{Altermagnetic spin-group symmetries} \;\;\;                                                                     & \textbf{Magnetic crystal structure} & \textbf{Non-relativistic band structure}                                                                                                                                            
 \\ \hline \hline \\
$\left[\bar{C}_{2} \| \mathcal{T}\right]$          &      Co-planar                     & Inversion symmetry       \\                                                                                                                                  \\ \hline \\
$[{\bf C}_\infty\parallel E]$                        & Collinear                                                    & Spin good quantum number \& {\bf k}-independent 
  \\                                                                                                                                                      \\ \hline \\
$\left[\bar{C}_{2} \| \mathcal{T}\right]\times[{C}_2\parallel A] \, [E\parallel{\bf H}]$            &         Compensated                &   Spin splitting \&  broken $\cal{T}$-symmetry      \\                                                                                                                                  \\ \hline \\
$\mathbf{H}$                       &                Sublattice spin-density anisotropy                            & Spin-Fermi-surface anisotropy         \\                                                                                                      \\ \hline \\
\textbf{L}$_{{\bf k}}\cap A{\bf H} \neq \emptyset$                                  &                             &  Spin-degenerate nodal lines or surfaces            \\                                                                       \\ \hline \\
\textbf{L}$_{{\bf k}}\cap A{\bf H} = \emptyset$                                &                                & Spin splitting  at crystal-momentum {\bf k}       \\                                                                                                                      \\ \hline \\
\textbf{L}$_{{\bf M}}\cap A{\bf H} = \emptyset$                          &                                       &  Spin splitting  at TRIM {\bf M}    \\                                                                                                                                                                                                             \\ \hline \\
Orbitally-degenerate $\boldsymbol\Gamma$-point\;\;\;  & &  Electric spin-splitting mechanism    \\
\end{tabular}
\caption{Spin-group symmetries and corresponding magnetic crystal structure and non-relativistic band structure characteristics. The first two lines regard symmetries of the spin-only group that apply to all three non-relativistic collinear phases. The remaining lines apply only to altermagnets. Symbol \textbf{L}$_{\bf k}$ marks the little group of real-space symmetry transformations which map crystal-momentum {\bf k} on itself or on a momentum which differs from {\bf k} by a reciprocal lattice vector. {\bf M} is a time-reversal invariant momentum (TRIM). Adapted from Ref.~ \cite{Smejkal2021a}}
\label{SLG}
\end{table}

\twocolumngrid

We conclude this section by summarizing the basic elements of the algorithm for determining the alternagnetic spin group (Eq.~(\ref{AMgroup})). It can be constructed by identifying:

\begin{enumerate}

\item
crystallographic group of the material,

\item
crystallographic group of the spin-sublattice (in the case of bipartite lattice the spin sublattice point group corresponds directly to Wyckoff position point group), 

\item
crystallographic rotation transformation connecting the opposite-spin sublattices.

\end{enumerate}

Taking RuO$_2$ as an example, the crystallographic point group is $4/mmm$, the sublattice (Wyckoff) point group is $mmm$, and the crystallographic point-group rotation transformation   connecting the opposite-spin sublattices is $C_{4z}$ (cf. Fig.~\ref{RuO2_symmetries}a).

\subsection{Identification rules}
Elementary rules for identifying the altermagnetic phase of a crystal can be summarized as follows:

\begin{enumerate}
\item
there is an even number of magnetic atoms in the unit cell and the number of atoms in the unit cell does not change between the non-magnetic and magnetic phases of the crystal (cf. two Ru atoms in RuO$_2$ unit cell, shown in Fig.~\ref{RuO2_symmetries}a),

\item
there is no inversion center between the sites occupied by the magnetic atoms from the opposite-spin sublattices (cf. the absence of the inversion center between the Ru$_{\rm A}$ and Ru$_{\rm B}$ sites  in RuO$_2$ because of the oxygen atoms, shown in Fig.~\ref{RuO2_symmetries}a),

\item
the two opposite-spin sublattices are connected by crystallographic rotation transformation, possibly combined with translation or inversion transformation (cf. the opposite-spin sublattices in RuO$_2$ connected by $C_{4z}{\bm t}$ transformation, shown in Fig.~\ref{RuO2_symmetries}a),

\item
the spin group is determined by the algorithm described in Sec.~II.B (cf. the RuO$_2$ spin group given by Eq.~(\ref{AMgroup}) with $A=C_{4z}$ and ${\bf H}=mmm$).
\end{enumerate}


\onecolumngrid

\bigskip

\begin{figure}[h!]
\begin{center}
\hspace*{-0cm}\epsfig{width=.8\columnwidth,angle=0,file=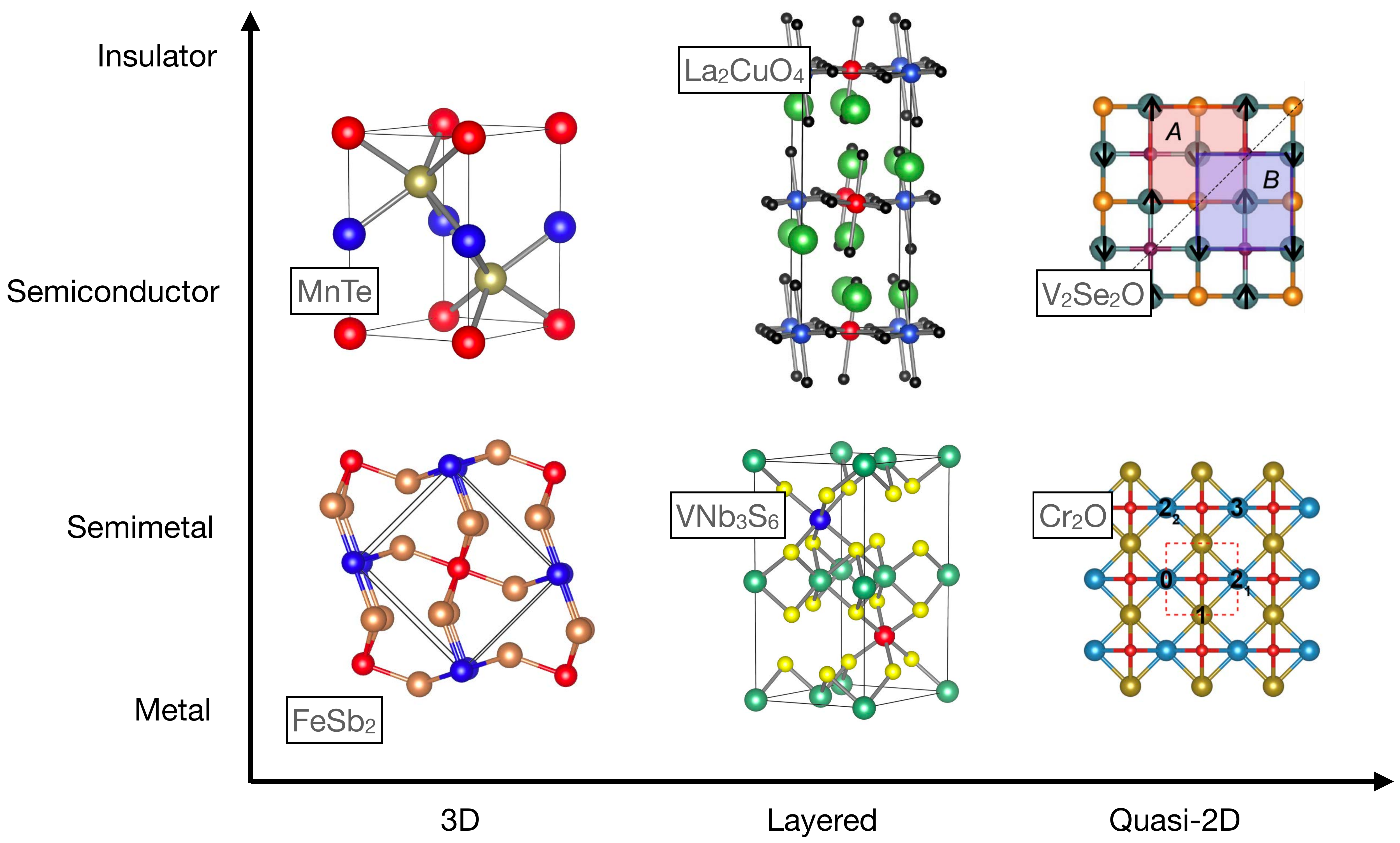}
\end{center}
\vspace{-.5cm}
\caption{Crystal structrures of selected altermagnetic candidates organized by dimensionality and conduction type. Adapted from Ref.~ \cite{Smejkal2021a,Mazin2021,Ma2021,Chen2021b}.
}
\label{fig_candidate_cryst}
\end{figure}

\twocolumngrid

\subsection{Material candidates}
The rules from Sec.~II.C can be used for high-throughput scanning of  altermagnetic material candidates. This section gives an overview of the predicted range of material types, illustrated on specific examples.

\begin{figure}[h!]
\begin{center}
\hspace*{-0cm}\epsfig{width=.62\columnwidth,angle=0,file=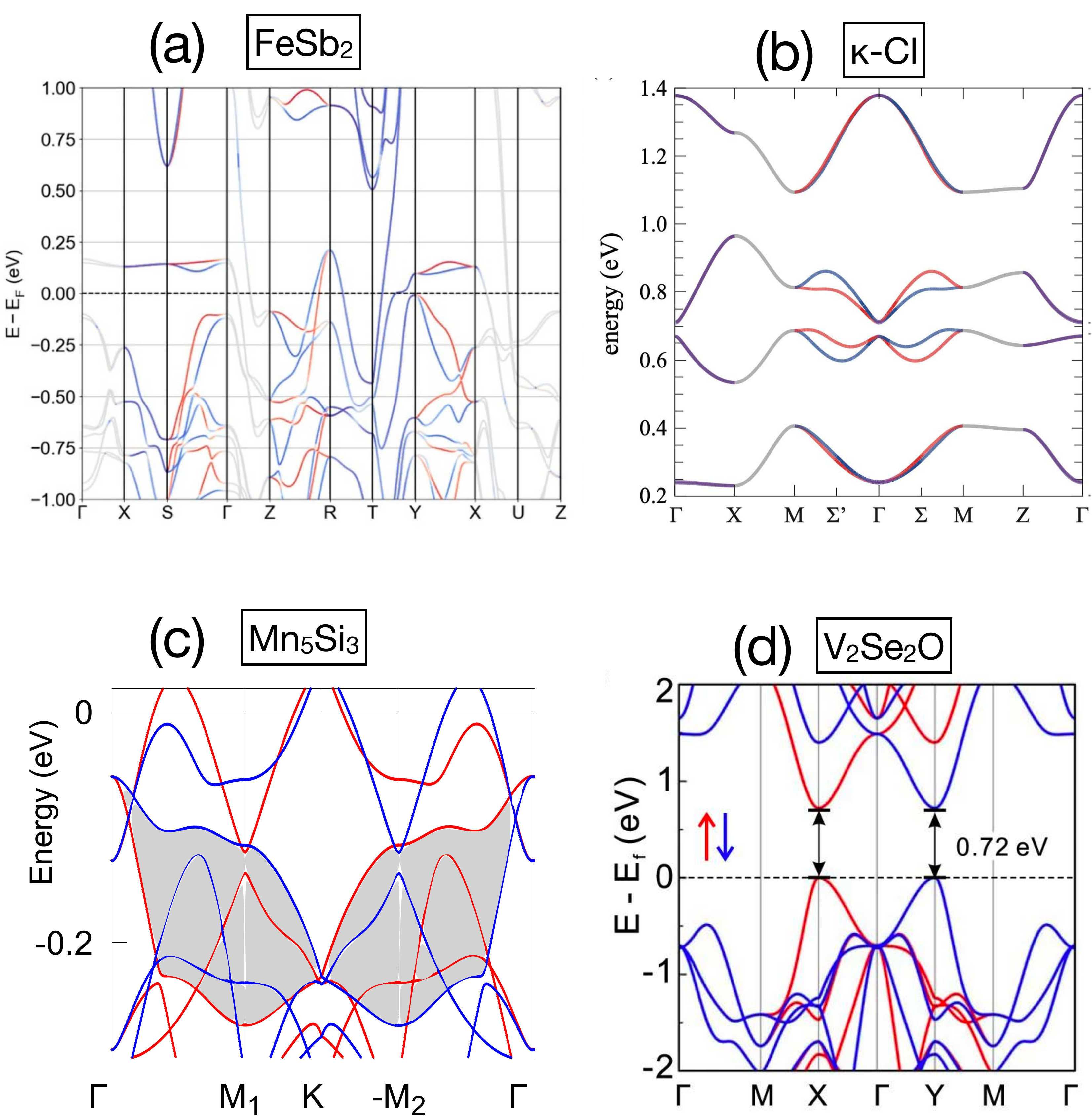}
\end{center}
\vspace{-.5cm}
\caption{(a-d) {\em Ab initio} spin-split band structures of depicted altermagnetic candidate materials. Adapted from Ref.~ \cite{Reichlova2020,Mazin2021,Ma2021,Seo2021}
}
\label{fig_candidate_band}
\end{figure}

\begin{figure}[h!]

\begin{center}
\hspace*{-0cm}\epsfig{width=.62\columnwidth,angle=0,file=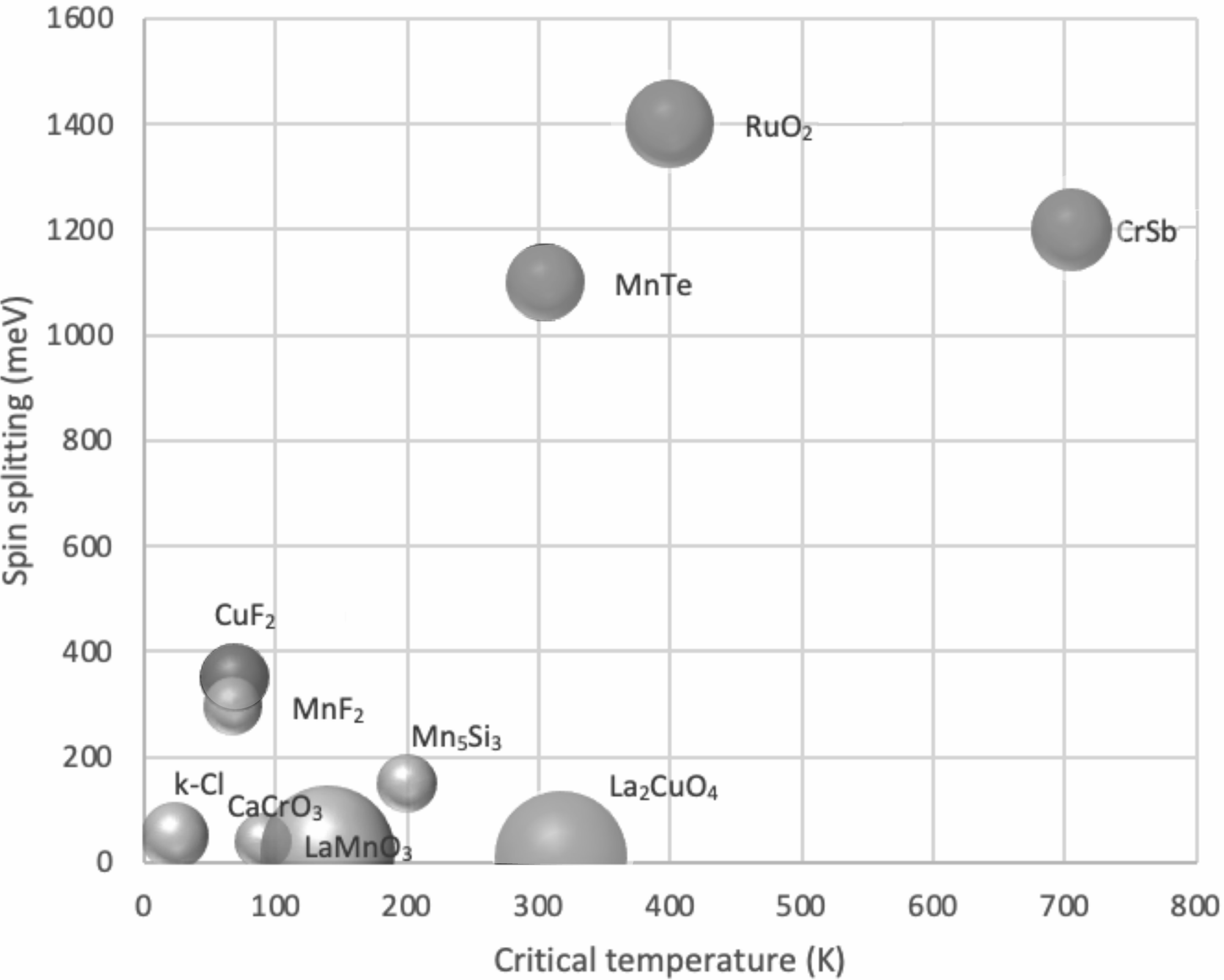}
\end{center}
\vspace{-.4cm}
\caption{Altermagnetic candidates identified from {\em ab initio} calculations, organizeed in a N\'eel temperature vs. altermagnetic spin-splitting strength diagram. Adapted from Ref.~ \cite{Smejkal2021a} and references therein. 
}
\label{fig_mater}
\end{figure}

Symmetry prohibits a realization of altermagnetism in one-dimensional (1D) chains, because of the absence of rotation transformations in 1D. On the other hand, Figs.~\ref{fig_candidate_cryst}-\ref{fig_mater}, and the list of material candidates given below illustrate that altermagnetism can occur in two-dimensional (2D) and three-dimensional (3D) crystals, the conduction types can cover the whole spectrum from insulators, semiconductors and semimetals, to metals and superconductors, and the structure and chemistry types can be also diverse: 
\begin{itemize}
\item quasi-2D oxide insulator V$_2$Se$_2$O \cite{Ma2021}  or semimetal Cr$_2$O \cite{Chen2021b},

\item 3D rutile fluoride or oxide insulators FeF$_2$ \cite{Lopez-Moreno2012}, MnF$_2$ \cite{Lopez-Moreno2016,Yuan2020}, MnO$_2$ \cite{Noda2016} and metal RuO$_2$ \cite{Smejkal2020,Ahn2019},

\item perovskite oxide insulators LaMnO$_3$ \cite{Okugawa2018,Yuan2021}, CaCrO$_3$ \cite{Naka2021} and parent cuprate of high-$T_c$ superconductor La$_2$CuO$_4$   \cite{Smejkal2021a},

\item ferrite insulator Fe$_2$O$_3$  \cite{Smejkal2021a},

\item pnictide with metal-insulator transition FeSb$_2$ \cite{Smejkal2021a,Mazin2021} and metal CrSb \cite{Smejkal2021a},

\item chalcogenide semiconductor MnTe \cite{Smejkal2021a} and (semi)metal VNb$_3$S$_6$ \cite{Smejkal2021a}, CoNb$_3$S$_6$ \cite{Smejkal2020},
  
\item silicide metal Mn$_5$Si$_3$ \cite{Reichlova2020},

\item organic insulator $\kappa$-Cl \cite{Naka2019}.
\end{itemize}
Crystallographic and spin groups,  and other characteristics of the selected altermagnetic material candidates are summarized in Tab.~\ref{table_mater}.  

A list of crystallographic symmetry groups that in principle allow for hosting the altermagnetic phase is given in Ref.~\cite{Smejkal2021a}. We also point out that altermagnetism can occur in structures with inversion symmetry (e.g. rutiles), or without inversion symmetry (e.g. VNb$_3$S$_6$ or CoNb$_3$S$_6$).  


\onecolumngrid

\newpage

\begin{table}[h!]
\begin{tabular}{@{}lllllllll@{}}
\toprule
                & Space group                       & Spin point group               & Conduction & T$_{\text{N}}$(K)          & Splitting(meV)             & Anisotropy      & Refs.                                         \\ \hline \hline
RuO$_2$    & $P4_2/mnm$          & $2_4/1_m 1_m 1_m$ & M   & 400            & 1400               & $d_{xy}$,P-2   & \cite{Smejkal2020,Ahn2019}    \\ \hline
KRu$_4$O$_8$           &  $I4/m$                             & $2_4/1_m$     & M   &                & 300                & $d_{x^{2}-y^{2}}$,  P-2 & \cite{Smejkal2021a}          \\ \hline
Mn$_5$Si$_3$  & $P6_3/mcm$                     & $2_m 2_m 1_m$    & M   & $\sim$200            & 150                & $d_{x^{2}-y^{2}}$, P-2     & \cite{Reichlova2020}         \\ \hline
(Cr,Fe)Sb$_2$            & $Pnma$                          & $2_m 2_m 1_m$    & M   &                & \textless{}200     & $d_{xy}$,P-2       & \cite{Mazin2021}             \\ \hline
CaCrO$_3$  & $Pnma$             &                                & M   & 90             & \textless{}200     & $d$-wave,P-2        & \cite{Naka2021}  \\ \hline
CrSb    & $P6_3/mmc$               & $2_6/2_m 2_m 1_m$          & M   & 705        & 1200               & $g$-wave,B-4            & \cite{Smejkal2021a}          \\ \hline
MnTe    & $P6_3/mmc$               & $2_6/2_m 2_m 1_m$          & I  & 310        & 1100               & $g$-wave,B-4            & \cite{Smejkal2021a}          \\ \hline
La$_2$CuO$_4$ & $Bmab$                           & $2_m 2_m 1_m$            & I  & \textless{}317 & 10                 & $d_{xy}$,P-2        &  \cite{Smejkal2021a} \\ \hline
MnO$_2$             & $P4_2/mnm$        & $2_4/1_m 1_m 1_m$  & I   &                & \textless{}900     & $d_{xy}$,P-2            & \cite{Noda2016}              \\ \hline
LaMnO$_3$  & $Pnma$              &                    & I   & 139,5          & 20                 & P-2            & \cite{Okugawa2018,Yuan2021}     \\ \hline
$\kappa$-Cl   & $Pnma$             &                                     & I   & 23             & \textless{}50      & $d_{xy}$,P-2         & \cite{Naka2019}              \\ \hline
MnF$_2$    &  $P4_2/mnm$        & $2_4/1_m 1_m 1_m$ & I   & 67             & 297                & $d_{xy}$,P-2          & \cite{Lopez-Moreno2016,Yuan2020}              \\ \hline
V$_2$Se$_2$O           &                  &                            & I   & -             & \textgreater{}1000 & $d_{x^{2}-y^{2}}$,P-2     & \cite{Ma2021}                \\ \hline
CuF$_2$    & $P2_1/c$                   & $2_2/2_m$              & I   & 69             & 350                & B-2            & \cite{Smejkal2021a}          \\ \bottomrule
\end{tabular}
\caption{Altermagnetic candidates identified from {\em ab initio} calculations. We list the non-magnetic space group,  spin point group, conduction type, transition temperature, and altermagnetic spin-splitting magnitude and anisotropy type.  "P/B--\#"  refers to the planar/bulk spin winding number, which around the   $\boldsymbol\Gamma$-point can take a value of 2, 4, or 6 (see Sec.~III).}
\label{table_mater}
\end{table}

\twocolumngrid

\section{Physical concepts}
To illustrate the potential and stimulate future research of altermagnetism in a broad condensed-matter physics field, we now discuss foreseen unique features of the altermagnetic phase in the context of several overarching physical concepts.

\subsection{Kramers theorem}

Energy bands are Kramers spin degenerate  \cite{Kramers1930,Wigner1932} across the whole Brillouin zone in all crystals that are invariant under the symmetry transformation that combines $\cal{T}$ and  space-inversion. Lifted Kramers spin-degeneracy by breaking this crystal symmetry brings forth a plethora of physically intriguing and technologically relevant phenomena, ranging from  topological phases of matter \cite{Franz2013,Bradlyn2017,Smejkal2018,Zang2018,Vergniory2019,Tokura2019,Xu2020,Elcoro2021}  and dissipationless Hall transport \cite{Nagaosa2010,Tokura2019,Smejkal2021b}, to charge-spin conversion effects in spintronic memory devices \cite{Chappert2007,Ralph2008,Bhatti2017,Manchon2019}.

\begin{figure}[h!]
\begin{center}
\hspace*{-0cm}\epsfig{width=.9\columnwidth,angle=0,file=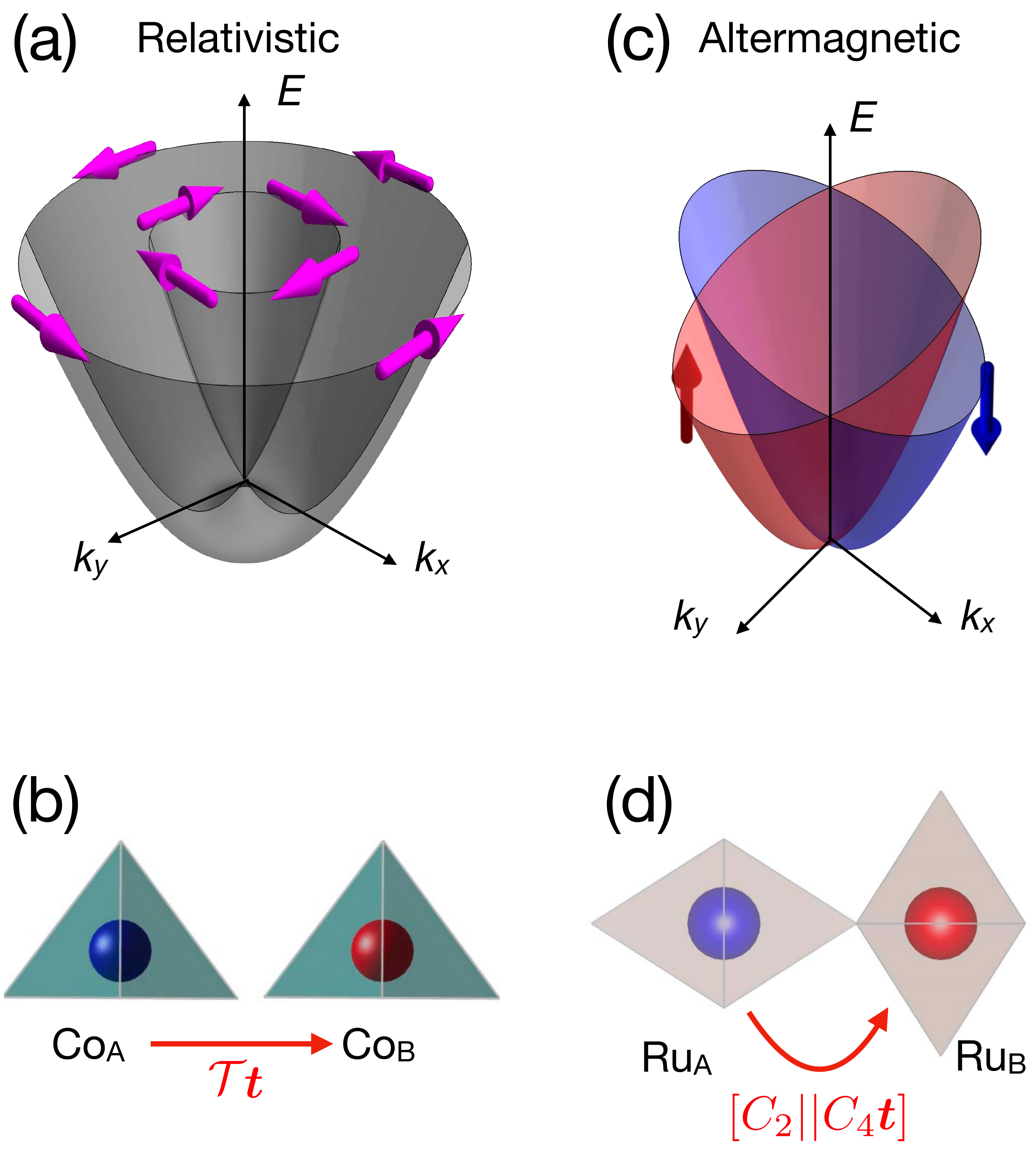}
\end{center}
\vspace{-.5cm}
\caption{ 
(a) Model relativistic Rashba spin-split bands. (b) Model of antiferromagnetic zero-magnetization crystal of BiCoO$_{\text{3}}$ with magnetic symmetry ${\cal T}{\bm t}$, and with broken space-inversion symmetry.  (c) Model non-relativistic altermagnetic spin splitting. (d) Model of altermnagnetic crystal of RuO$_2$ with non-relativistic spin symmetry $[C_2\parallel C_4{\bm t}]$.  Adapted from Ref.~ \cite{Yamauchi2019,Smejkal2021a}.}
\label{relativistic}
\end{figure}

For the many decades of spin-physics research, lifting of the Kramers spin degeneracy in energy bands has been considered to originate from two basic mechanisms. The first one links the broken space-inversion symmetry to the spin space by the electron's relativistic spin-orbit coupling \cite{Winkler2003,Armitage2018}.  It results in inversion-asymmetric spin-split energy bands, and  non-collinear spin textures (e.g. Rashba) in the momentum space, illustrated in Fig.~\ref{relativistic}a. 

The second mechanism is associated with $\cal{T}$-symmetry breaking by external magnetic field or by internal magnetization of ferromagnets (ferrimagnets) \cite{Landau1984}. Microscopically, the latter tends to be dominated by a non-relativistic magnetic-exchange interaction, and is commonly modelled by a momentum-independent effective Zeeman term, as illustrated in Fig.~\ref{FM_AF_AM}a.

All magnetically ordered crystals have broken $\cal{T}$-symmetry. While this directly leads to the effective Zeeman spin splitting of energy bands in ferromagnets, spin splitting   has been commonly considered to be excluded in crystals with a compensating antiparallel arrangement of magnetic moments  \cite{Turov1965,Neel1971,Nunez2006,Surgers2014,Surgers2016,Ghimire2018}. Indeed, there are two types of Kramers spin-degenerate antiferromagnets. 

The first type has  a symmetry combining $\cal{T}$ with translation $\bm t$  in the antiferromagnetic crystal. The ${\cal T}\bm t$-symmetry defines type-IV magnetic space groups. Among those, only the antiferromagnetic crystals with space-inversion symmetry have the Kramers spin-degenerate bands. As a result, the bands have $\cal{T}$-symmetry, $\epsilon(s,{\bf k})=\epsilon(-s,-{\bf k})$,  and inversion symmetry, $\epsilon(s,{\bf k})=\epsilon(s,-{\bf k})$, apart from being spin degenerate, $\epsilon(s,{\bf k})=\epsilon(-s,{\bf k})$. Examples  are FeRh or MnBi$_2$Te$_4$ \cite{Marti2014,Otrokov2019}. In the non-relativistic limit and for collinear antiferromagnetic order (cf. Sec.~II.B), the Kramers spin degeneracy is protected by the spin-group symmetry $[{C}_2\parallel {\bm t}]$  alone, i.e., independent of whether the antiferromagnetic crystal is or is not inversion symmetric. We note that the non-relativistic collinear symmetry $[{C}_2\parallel {\bm t}]$ does not imply that all materials described by type-IV magnetic space groups 
have necessarily vanishing spin splitting   in the non-relativistic limit.  This is because type-IV magnetic space groups encompass also non-collinear magnets. However, all materials from type-IV magnetic space groups have $\cal{T}$-symmetric bands, whether or not relativistic effects are included.

The second type of antiferromagnetic crystals with Kramers spin-degenerate bands break space-inversion and $\cal{T}$ (or ${\cal T}{\bm t}$)  symmetries on their own, but have a symmetry combining the two transformations.  In this case, the Kramers spin-degenerate bands have broken  inversion symmetry and broken $\cal{T}$-symmetry. Here CuMnAs or Mn$_2$Au are among the prominent material examples  \cite{Zelezny2014,Wadley2016,Smejkal2017c,Elmers2020,Fedchenko2021}. The Kramers spin-degeneracy of non-relativistic  bands in these collinear antiferromagnets is protected by spin symmetry $[{C}_2\parallel \bar{E}]$, in combination  with the spin-only group symmetry $\left[\bar{C}_{2} \| \mathcal{T}\right]$ (cf. 1$^{\rm st}$ line in Tab.~\ref{SLG}). In addition, the $[{C}_2\parallel \bar{E}]$-symmetry protects $\cal{T}$-symmetry of the non-relativistic bands, and $\left[\bar{C}_{2} \| \mathcal{T}\right]$ protects inversion symmetry of bands in the non-relativistic limit. Breaking of inversion symmetry and $\cal{T}$-symmetry in the band structure of this type of collinear antiferromagnets is, therefore, purely of relativistic origin.

While magnets with the compensated magnetic order were commonly associated with spin-degenerate bands,  Zeeman or relativistic spin-splitting mechanisms were also discussed in antiferromagnets. An effective ferromagnetic-like Zeeman splitting, and the corresponding  $\cal{T}$-symmetry breaking in the band structure, was associated in antiferromagnets with a net moment, suggested to occur due to canting, for instance in GdPtBi  \cite{Suzuki2016}. Alternatively, a real Zeeman spin splitting was considered in antiferromagnets due to an external magnetic field   \cite{Ramazashvili2008,Ramazashvili2009,Rozbicki2011}.  A $\cal{T}$-symmetric relativistic Rashba splitting was predicted in antiferromagnets such as BiCoO$_{\text{3}}$ with broken space-inversion symmetry, and with the opposite Co sublattices connected by the ${\cal T}{\bm t}$-symmetry, as shown in Figs.~\ref{relativistic}a,b \cite{Yamauchi2019}.  Another example of an unconventional magnetic and relativistic splitting has been experimentally demonstrated in surface states of antiferromagnetic NdBi  \cite{Schrunk2022}. Both of these types of spin splitting offer intriguing interplay with antiferromagnetism. However, they also inherit a net magnetization (Fig.~\ref{FM_AF_AM}a) or a non-collinear spin texture (Fig~\ref{relativistic}a) in the band structure, characteristic of conventional ferromagnets or relativistic spin-orbit-coupled systems, respectively. 

Lifting of the Kramers spin degeneracy by the altermagnetic phase, illustrated in Figs.~\ref{relativistic}c,d,  is distinct from the conventional mechanisms. Unlike the relativistic spin-orbit coupling mechanism, it does not require breaking of the space-inversion symmetry. In fact, the non-relativistic band structure of altermagnets has the inversion symmetry protected by the spin-group symmetry corresponding to the co-planarity of the magnetic order, as shown on the 1$^{\rm st}$ line in Tab.~\ref{SLG}. This applies independently of the presence or absence of inversion symmetry in the magnetic crystal structure \cite{Smejkal2021a}. Also unlike the relativistic spin-orbit coupling mechanism, the non-relativistic electronic states in the altermagnetic bands have a common spin-quantization axis and spin is a good quantum number. These characteristics are protected  by the spin-group symmetry corresponding to the collinearity  of the magnetic order, as highlighted on the 2$^{\rm nd}$ line in Tab.~\ref{SLG}. 

Comparing to ferromagnets, altermagnets share  the strong non-relativistic $\cal{T}$-symmetry breaking and spin splitting in the band structure. In altermagnets, these characteristics are allowed by the spin-group symmetry shown on the 3${\rm rd}$ line in Tab.~\ref{SLG}. The distinction from ferromagnets is that the same  spin-group symmetry also protects the zero non-relativistic net magnetization in altermagnets.

\subsection{Fermi-liquid instabilities}

In Fermi-liquid theory, interactions among electron  quasiparticles are described by Landau parameters in spin-singlet and spin-triplet channels, using an orbital angular momentum partial-wave expansion. Large (negative) values of Landau parameters lead to Pomeranchuk Fermi-liquid instabilities  \cite{Pomeranchuk1959,Wu2007}. A prominent  example of an isotropic $s$-wave  instability in the spin-triplet channel is Stoner ferromagnetism, which corresponds to the momentum-independent effective Zeeman spin splitting in the electronic band structure. 

Theoretically, a rich landscape of quantum ordered phases is linked to  anisotropic (non-zero angular momentum) Landau parameters  \cite{Wu2007}. However, experimental indications of anisotropic Fermi-liquid instabilities are rare.  An example are nematic-phase instabilities in the spin-singlet channel with non-zero angular momenta. Their typical characteristics are anisotropic distortions of Fermi surfaces.  Nematic instabilities were considered in fractional quantum Hall systems, Mott insulators or high-$T_c$ superconductors -- all belonging to the family of complex strongly correlated systems \cite{Wu2007}. 

In analogy to Stoner ferromagnetism, non-zero angular momentum instabilities in the spin-triplet channel typically break the $SU$(2) symmetry of the non-relativistic non-magnetic Fermi liquid \cite{Wu2007}. A characteristic feature of a spin-triplet  $p$-wave instability is a shift of spin-up and spin-down Fermi surfaces in opposite directions. Unlike Stoner ferromagnetism dominated by the exchange interaction, and in analogy to the nematic phases, the spin-triplet  $p$-wave instability  was considered in correlated systems, e.g., in heavy fermion compounds \cite{Wu2007}.

Altermagnetism is not described by a $p$-wave (or higher odd-parity wave) instability in the spin-triplet channel. This is seen from the spin-group symmetry on the 1$^{\rm st}$ line in Tab.~\ref{SLG}, reflecting the co-planarity of the magnetic crystal order, and the corresponding inversion symmetry of the spin-split band structure. On the other hand, altermagnetism can be associated with even-parity wave Fermi liquid instabilities in the spin-triplet channel \cite{Ahn2019,Smejkal2021a}. We now illustrate on the band structure of RuO$_2$ that the predicted characteristics of the altermagnetic Fermi-liquid instabilities are extraordinary.

\begin{figure}[h!]
\vspace{.5cm}
\begin{center}
\hspace*{0cm}\epsfig{width=1\columnwidth,angle=0,file=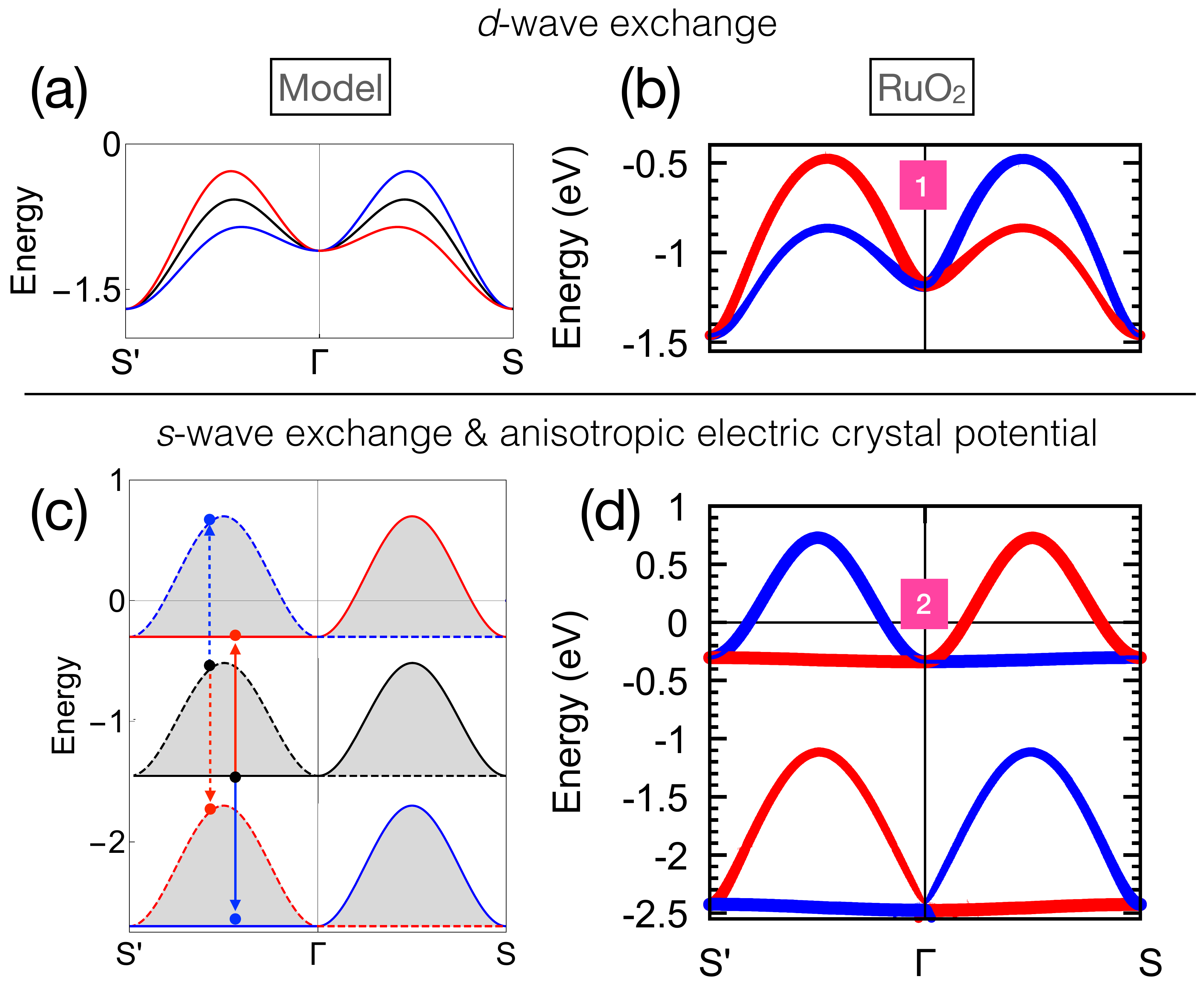}
\end{center}
\vspace{-.5cm}
\caption{(a) Schematic diagram of the anisotropic ($d$-wave) exchange Fermi-liquid instability in the altermagnet. Black line corresponds to the spin-degenerate band in the non-magnetic phase while red and blue lines are spin-split bands in the altermagnetic phase. (b) Spin-projected (and orbital-projected) {\em ab initio} bands of RuO$_{\text{2}}$ in the energy window corresponding to (a). (c)  Schematic diagram of the isotropic ($s$-wave) exchange Fermi-liquid instability combined with anisotropic electric crystal potential in the altermagnet. Solid and dashed black lines correspond to the spin-degenerate bands dominated by one or the other sublattice in the non-magnetic phase, respectively. Red and blue lines are spin-split bands in the altermagnetic phase.  (d) Spin-projected (and orbital-projected) {\em ab initio} bands of RuO$_{\text{2}}$ in the energy window corresponding to (c). Pink boxes with labels "1" and "2" correspond to the full {\em ab initio} bands of RuO$_{\text{2}}$ shown in Fig.~\ref{ab_initio_AM}c. Adapted from Ref.~\cite{Smejkal2021a}.
} 
\label{spin-splitting-mechanisms}
\end{figure}

A spin-splitting mechanism   due to an anisotropic exchange interaction \cite{Yuan2020,Ahn2019,Hayami2020,Smejkal2022} can be identified in parts of the RuO$_2$ band structure with a single two-fold spin-degenerate band in the non-magnetic phase, which undergoes in the altermagnetic phase an anisotropic momentum-dependent spin splitting with alternating sign \cite{Smejkal2021a}. This is illustrated in Figs.~\ref{spin-splitting-mechanisms}a,b on a schematic diagram and {\em ab initio} bands of RuO$_2$. Remarkably, the mechanism is dominated by anisotropic exchange interaction, i.e., it persists   without including correlation effects beyond the local-spin-density approximation \cite{Ahn2019,Smejkal2021a}.

\begin{figure}[h!]
\begin{center}
\hspace*{-0cm}\epsfig{width=.65\columnwidth,angle=0,file=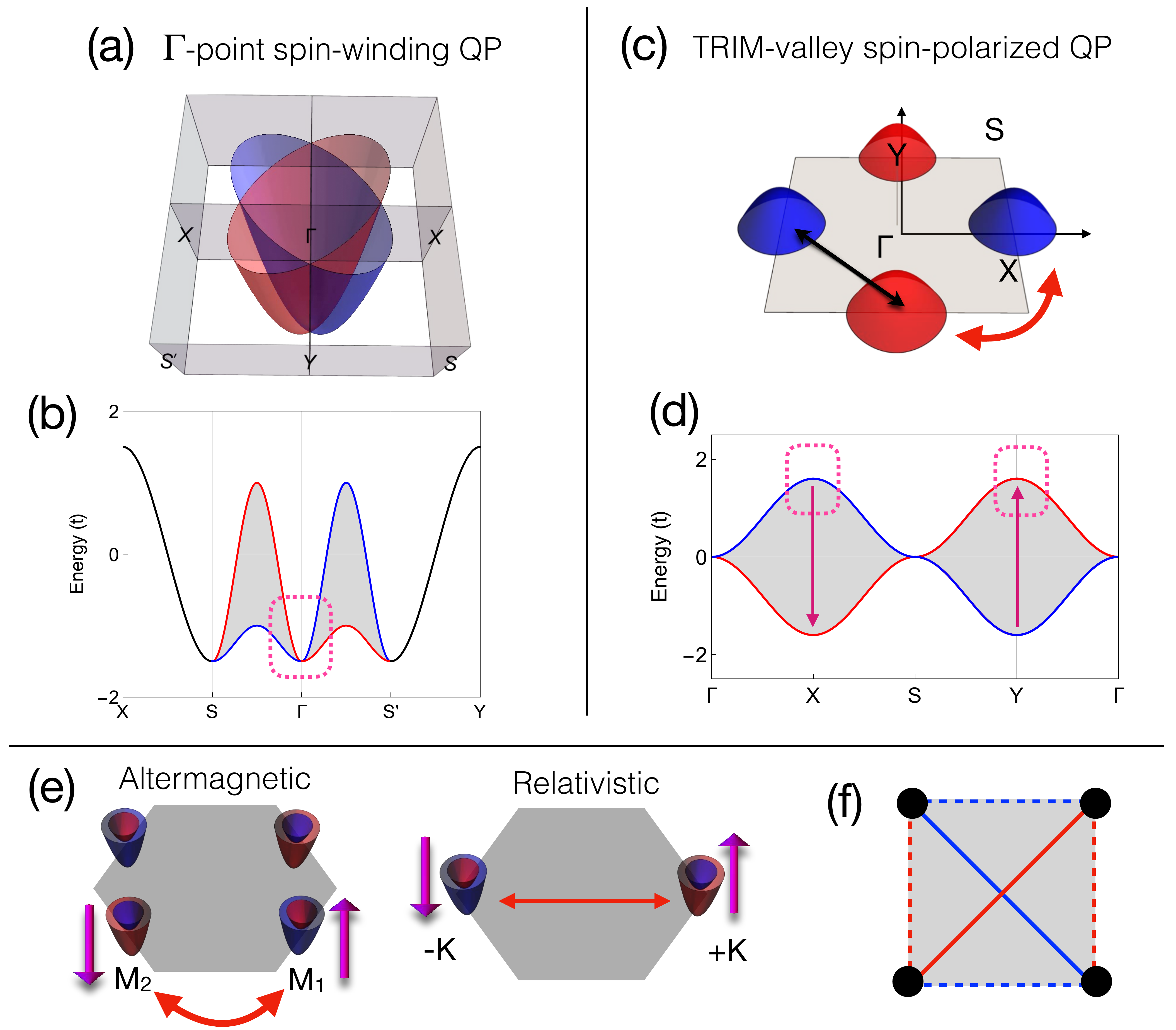}
\end{center}
\vspace{-.5cm}
\caption{(a,b) 
Model of the altermagnetic quasiparticle with quadratic dispersion around the spin-degenerate $\mathbf\Gamma$-point, and spin-winding number 2 around the $\mathbf\Gamma$-point. The model corresponds to Eq.~(\ref{xy}).
(c,d)  Model of the altermagnetic spin-polarized valley quasiparticle with no spin winding in the valley. The model corresponds to Eq.~(\ref{x2y2}). The center of a valley is at TRIM, and the spin polarization is opposite at TRIM ${\bf X}$ and ${\bf Y}$. (e) Schematic illustration of distinct symmetries of band structures with non-relativistic altermagnetic and relativistic non-magnetic valleys. (f) Real-space spin-dependent hoppings used to construct model band structures in (a-c). Adapted from Refs.~\cite{Mazin2021,Reichlova2020,Smejkal2022}.
} 
\label{QP}
\end{figure}

An effective single-particle two-band Hamiltonian that models this mechanism contains, apart from the 
common kinetic-energy hopping term, 
a spin-dependent hopping due to the anisotropic exchange interaction in the altermagnetic state  \cite{Reichlova2020,Smejkal2022,Mazin2021}:
\begin{equation}
H=2t\cos {k_{x}}\cos {k_{y}} + 2t_{J}\sin{k_{x}}\sin {k_{y}}\sigma_{z}.
\label{xy}
\end{equation}
The model  band structure is plotted in Figs~\ref{QP}a,b. 
The energy spectrum exhibits spin-degenerate nodal surfaces at $k_{x,y}=0,\pi$, marked in grey in Fig.~\ref{QP}a, and protected by mirror plane symmetries that transform one spin sublattice on the opposite-spin subblatice, and are contained in the little groups of the nodal-surface momenta (cf. 5$^{\rm th}$ line in Tab.~\ref{SLG}). The resulting nodal structure and spin splitting modulation pattern corresponds to a $d_{xy}$-wave symmetry.   The  characteristic $d_{xy}$-wave spin-up and spin-down Fermi surfaces are anisotropic and mutually rotated by  90$^\circ$, following the $[{C}_2\parallel C_{4}]$ spin-group symmetry. {\em Ab initio} bands of RuO$_2$ and the corresponding model thus illustrate an extraordinary spin-triplet  Fermi-liquid instability  that, unlike the $s$-wave Stoner ferromagnetism, is anisotropic and, unlike the $p$-wave heavy-fermion compounds, is uncorrelated and occurs in the even-parity $d$-wave channel.

Remarkably, the {\em ab initio} band structure of RuO$_2$ demonstrates an additional distinct spin-splitting mechanism. In this case, the size and momentum-dependence of a strong non-relativistic altermagnetic spin splitting is determined by  the band splitting due to an anisotropic crystal potential of the non-magnetic phase \cite{Smejkal2021a}. 
This unconventional electric spin-splitting mechanism is illustrated  on a schematic diagram in Fig.~\ref{spin-splitting-mechanisms}c. In the non-magnetic state, there are two spin-degenerate bands that cross at the four-fold spin and orbital-degenerate $\boldsymbol\Gamma$-point, while the orbital degeneracy is lifted away from the   $\boldsymbol\Gamma$-point (cf. 8$^{\rm th}$ line in Tab.~\ref{SLG}). One of the two spin-degenerate bands has a dominant projection on one sublattice while the other band on the other sublattice. The bands are anisotropic  due to the anisotropy of the electric crystal potential (cf. 4$^{\rm th}$ line in Tab.~\ref{SLG}). The anisotropies of the spin-degenerate bands corresponding to the two sublattices are mutually rotated by 90$^\circ$, reflecting the real-space $C_4$ rotation symmetry which transforms one crystal sublattice on the other. As a result, there is a mutual momentum-dependent splitting between the two spin-degenerate bands away from the $\boldsymbol\Gamma$-point. 

The altermagnetic phase brings  an additional momentum-independent (isotropic) exchange interaction, with opposite sign in the bands corresponding to opposite-spin sublattices. As a result, two pairs of spin-split bands form with opposite sign of the spin splitting. For a given pair, the size and momentum-dependence of the spin splitting  is a copy of the size and momentum-dependence of the orbital splitting in the non-magnetic state. The presence of this microscopic spin-splitting mechanism in RuO$_2$ is again confirmed by {\em ab initio} calculations shown in Fig.~\ref{spin-splitting-mechanisms}d. In this case, the altermagnet can be viewed as two interpenetrating $s$-wave Stoner ferromagnets with opposite magnetizations that, due to the interplay with anisotropies of the electric crystal potential, generate spin-split $d$-wave-like Fermi surfaces.

The potential richness of the landscape of altermagnetic Fermi-liquid instabilities can be further inferred by inspecting the symmetries of all altermagnetic spin groups. Each altermagnetic spin point group can be associated with a given minimum even-parity wave anisotropy of the spin-split Fermi surfaces near the $\boldsymbol\Gamma$-point   \cite{Smejkal2021a}. Apart from $d$-wave, this minimum anisotropy can have a $g$-wave or $i$-wave form \cite{Hayami2020,Smejkal2021a}.

\onecolumngrid

\bigskip

\twocolumngrid

\subsection{Electron and magnon quasiparticles}
The predicted extraordinary Fermi-liquid instabilities in altermagnets can generate a variety of unconventional electron quasiparticles. An example can be illustrated on the energy bands of the model two-band Hamiltonian (\ref{xy}) in the ${\bf k}\cdot{\bf p}$ approximation around the  $\boldsymbol\Gamma$-point, highlighted by a dashed rectangle in Fig.~\ref{QP}b. The spin-dependent part of the band structure is given by $2t_Jk_{x}k_{y}\sigma_{z}$. The spin degeneracy of the $\boldsymbol\Gamma$-point is generally protected in altermagnets by the spin-group symmetry $[{C}_2\parallel A]$, because the $\boldsymbol\Gamma$-point is invariant under any real-space symmetry transformation (including the rotations $A$). The  $\boldsymbol\Gamma$-point spin degeneracy is analogous to the $\cal{T}$-symmetric relativistic bands. Here an example is the Rashba model whose spin-dependent part is given by  $\lambda \left( k_{x}\sigma_{y} - k_{y} \sigma_{x}  \right)$ (Fig.~\ref{relativistic}a). However, unlike the linearly dispersing quasiparticles around the $\boldsymbol\Gamma$-point of the inversion-asymmetric relativistic bands, the altermagnetic quasiparticles in the above model have  a quadratic dispersion around the $\boldsymbol\Gamma$-point, in line with the  general inversion symmetry of  bands in altermagnets (cf. 1$^{\rm st}$ line in Tab.~\ref{SLG}).   

The altermagnetic quasiparticles are spin polarized away from the $\boldsymbol\Gamma$-point  and can be assigned a spin-winding number. In analogy to relativistic systems, the spin-winding number describes how many times spin reverses when completing a closed path around the $\boldsymbol\Gamma$-point. However, the spin-winding number of the above altermagnetic quasiparticle model is 2, in contrast to the spin-winding number 1 in the relativistic Rashba model.  The spin-winding numbers around the $\boldsymbol\Gamma$-point in altermagnets have even-integer values, again because of the inversion symmetry of the bands. The spin winding of the altermagnetic quasiparticles can be planar (cf. relativistic Rashba or Dresselhaus spin-textures) or bulk (cf. relativistic Weyl spin-texture) \cite{Smejkal2021a}. The allowed even-integer values are 2, 4 or 6 in the vicinity of  the $\boldsymbol\Gamma$-point, and are determined by the spin-group of the altermagnet \cite{Smejkal2021a}. This illustrates to potential richness of the spin-polarized quasiparticles in altermagnets around  the $\boldsymbol\Gamma$-point.

We point out that the spin-winding number in the relativistic systems is associated with continuously varying spin direction in the momentum space. In contrast, altermagnets show that non-zero integer invariants, describing how many times the quasiparticle spin reverses when completing a closed path around the $\boldsymbol\Gamma$-point, can exist also in systems where all spins share a common spin quantization axis, and spin is a good quantum number.

A different type of predicted altermagnetic electron quasiparticles can be illustrated on a two-band model Hamiltonian \cite{Reichlova2020,Smejkal2022,Mazin2021}:
\begin{equation}
H=\pm 2t_{J}\left( \cos {k_{x}}- \cos {k_{y}} \right)\sigma_{z}, 
\label{x2y2}
\end{equation}
whose energy spectrum is shown  in Fig.~\ref{QP}d. 
(Around the $\Gamma$-point, Eq.~\eqref{x2y2}  is  related to the model in Eq.~\eqref{xy} by a 45$^\circ$ rotation of the momentum space, and by setting $t=0$).

In the ${\bf k}\cdot{\bf p}$ approximation around time-reversal invariant momenta (TRIMs) {\bf X} and {\bf Y}, highlighted by dashed rectangles in Fig.~\ref{QP}d, the spectrum takes  a form of spin-split valleys  given by (see Figs.~\ref{QP}c,d), 
\begin{eqnarray}
E_{\pm}({\bf X},\textbf{k})&=&\pm t_J(4-k^2), \nonumber \\
E_{\pm}({\bf Y},\textbf{k})&=&\mp t_J(4-k^2).
\label{TMR-model}
\end{eqnarray} 
(Recall that   a momentum {\bf k} is time-reversal invariant when it differs from $-{\bf k}$ by a reciprocal lattice vector.)   

The spin-group symmetry condition allowing for the spin-split TRIMs is given on the 7$^{\rm th}$ line in Tab.~\ref{SLG}. The possibility to observe  spin-split valleys around TRIMs in real materials is predicted by {\em ab initio} band structure calculations of the altermagnetic phase in, e.g.,  Mn$_5$Si$_3$  \cite{Reichlova2020} (Fig.~\ref{fig_candidate_band}c).  Besides 3D crystals, the altermagnetically spin-split valleys can also form in 2D materials  \cite{Egorov2021,Egorov2021a,Ma2021}, as predicted, e.g., in {\em ab initio} calculations of the band structure of a monolayer insulator V$_2$Se$_2$O \cite{Ma2021} (Fig.~\ref{fig_candidate_band}d).  

Locally, the individual valleys around the ${\bf X}$  and ${\bf Y}$ TRIMs are isotropic in the above model. This illustrates that altermagnets can host spin-polarized quasiparticles analogous to the model non-relativistic $s$-wave Stoner ferromagnet, with  no spin winding around the TRIM. However, unlike ferromagnets, the altermagnetic spin-group symmetries impose that each spin-split TRIM has a counterpart TRIM elsewhere in the Brillouin zone with opposite spin splitting. The presence of these TRIM pairs is protected by the $[{C}_2\parallel A] \, [E\parallel{\bf H}]$ spin-group symmetries (cf. 3$^{\rm rd}$ line in Tab.~\ref{SLG}). 

The altermagnetic spin-polarized quasiparticles in separate local valleys in the momentum space are reminiscent of relativistic spin-polarized valley-quasiparticles in non-magnetic hexagonal 2D materials, such as  transition metal dichalcogenides \cite{Schaibley2016}. The common features shared by the altermagnetic and relativistic quasiparticles are the opposite spin polarization in valleys occupying different parts of the Brillouin zone, and the zero net spin polarization when integrated over the whole Brillouin zone. However, only altermagnets allow these valleys to be centered at TRIMs, as highlighted in Fig.~\ref{QP}e. In the non-magnetic relativistic systems, spin splitting is excluded  by $\cal{T}$-symmetry not only at the $\boldsymbol\Gamma$-point, but at all TRIMs. 

So far we discussed the electron quasiparticles from the symmetry perspective limited to the spin-group transformations acting on the spin and momentum-dependent band structure. Additional rich quasiparticle physics, including higher-order degeneracy quasiparticles, can emerge from the analysis of spin-group transformations acting on the electron wavefunctions (spin-group representations) \cite{Smejkal2021a,Corticelli2022,Liu2021,Lenggenhager2022}.

Besides the electron energy spectra and quasiparticles, we foresee that the real-space symmetries of the altermagnetic crystal structure will be also reflected in unique characteristics of the spin-wave spectra and magnon quasiparticles \cite{Marmodoro2022_am_magnons}. 

The typically leading contribution to the magnon spectra can be obtained by mapping the spin-dependent electronic structure on the Heisenberg Hamiltonian, $H=-\sum_{i\neq j}J_{ij}\hat{{\bf e}}_i\hat{{\bf e}}_j$ \cite{Halilov1998,Corticelli2022,Marmodoro2022}. Here $\hat{{\bf e}}_i$ is the direction of the magnetic moment around
an atom at position $R_i$, and  $J_{ij}$ are  Heisenberg exchange coupling parameters. In the Heisenberg model, the
real and spin space transformations are decoupled and  the symmetries of the corresponding magnon bands can be described by the non-relativistic spin-group formalism \cite{Brinkman1966,Corticelli2022}. 

In antiferromagnets, translation or inversion symmetry transformations connecting  opposite-spin sublattices protect double-degeneracy of the magnon spectra \cite{Brinkman1966,Corticelli2022}. This has been commonly illustrated on the opposite-spin sublattices of rutile crystals, while omitting the presence of non-magnetic  atoms  in these crystals \cite{Brinkman1966,Rezende2019,Corticelli2022}. We have seen in Sec.~II, however, that the non-magnetic O-atoms in rutile RuO$_2$ (or F-atoms in rutile FeF$_2$ and MnF$_2$) break the translation and inversion symmetries connecting the opposite-spin sublattices. Instead of classic antiferromagnets  \cite{Neel1953,Brinkman1966,Rezende2019,Corticelli2022},  rutiles are  the prototypical representatives of altermagnetism, with interlinked unique properties of the real-space crystal structure and momentum-space electronic and magnonic band structures \cite{Marmodoro2022_am_magnons}.

In analogy to the electronic band structure, the double-degeneracy of  magnon bands is predicted to be lifted in altermagnets, with the sign of the splitting alternating across the magnon Brillouin zone \cite{Marmodoro2022_am_magnons}. Magnons in altermagnets can also feature  antiferromagnetic-ferromagnetic dichotomy. The dispersion of altermagnetic magnons can be linear around the degenerate $\boldsymbol\Gamma$-point, in analogy to antiferromagnets, while the Landau damping  in altermagnets can be comparably  low to the damping of  magnons in ferromagnets due to the comparably large spin splitting of the electron quasiparticles  \cite{Marmodoro2022,Marmodoro2022_am_magnons}.

\subsection{Berry phase and non-dissipative transport}

Berry phase  is a general concept in quantum mechanics \cite{Berry1984a}.  A prototypical example is the Aharonov-Bohm phase given by a real-space path integral of the electrodynamic vector potential along a closed loop or, equivalently, by an integral of the magnetic field over an area enclosed by the loop. The phase can be macroscopically observable by resistance oscillations in an applied magnetic field. 

In the crystal momentum space, a Berry connection analogue of the electrodynamic vector potential, and a Berry curvature analogue of the magnetic field, 
\begin{eqnarray}
\bm{\mathcal{A}}_n({\bf k})&=&i\langle u_{n{\bf k}}|\nabla_{\bf k}u_{n{\bf k}}\rangle \nonumber \\
\bm{\mathcal{B}}_n({\bf k})&=&\nabla_{\bf k}\times\bm{\mathcal{A}}_n({\bf k})  ,
\label{Berry-con-cur}
\end{eqnarray}
can also generate macroscopic observables. A prominent example is the non-dissipative Hall current given by the transverse conductivity \cite{Nagaosa2010}, 
\begin{equation}
\sigma^{\rm Hall}_{xy}=-\frac{e^2}{\hbar}\sum_{n}\int_{\rm BZ} \frac{d^3k}{(2\pi)^3} f[\varepsilon_n({\bf k})] {\mathcal{B}}_n^z({\bf k}) .
\label{Berry}
\end{equation} 
Here $f[\varepsilon_n({\bf k})]$ is the Fermi-Dirac distribution function, $\varepsilon_n({\bf k})$ is the energy of the  Bloch state in band $n$ with crystal momentum {\bf k}, and $u_{n{\bf k}}({\bf r})$ is the periodic part of the Bloch wavefunction.

Altermagnets are predicted to bring unique elements into the physics of Berry phase phenomena \cite{Smejkal2021b,Smejkal2020,Feng2020a,Reichlova2020,Samanta2020,Naka2020,Mazin2021}. The Berry curvature near the $\boldsymbol\Gamma$-point of a ${\bf k}\cdot{\bf p}$ altermagnet-Rashba model \cite{Smejkal2021b},  $tk^2+2t_Jk_{x}k_{y}\sigma_{z}+\lambda \left( k_{x}\sigma_{y} - k_{y} \sigma_{x}  \right)$, is given by
\begin{equation}
\mathcal{B}(k)_{\pm}=\mp\frac{2t_J\lambda^{2}k_{x}k_{y}}{\sqrt{4t_J^{2}(k_{x}k_{y})^{2}+\lambda^{2}k^{2}}}. 
\label{AM-Berry}
\end{equation} 
Eq.~(\ref{AM-Berry}) illustrates that the characteristic even-parity wave ($d$-wave) anisotropy in the non-relativistic band structure of altermagnets can be  also reflected in their relativistic Berry curvature. In contrast, a counterpart ferromagnet-Rashba model,  $tk^2+\Delta \sigma_{z}+\lambda \left( k_{x}\sigma_{y} - k_{y} \sigma_{x}  \right)$, gives an isotropic Berry curvature near the $\boldsymbol\Gamma$-point \cite{Culcer2003,Nunner2007,Dugaev2008,Nagaosa2010,Xiao2010b,Smejkal2021b}, reflecting the principally isotropic $s$-wave nature of ferromagnetism.

The Berry curvature tends to reach the highest values near band (anti)crossings  \cite{Nagaosa2010,Smejkal2021b}, which implies another outstanding feature of altermagnets. In contrast to the typically accidental (anti)crossings in ferromagnets, the spin-group symmetries of altermagnets impose the presence of the nodal lines or nodal surfaces in the band structure (cf. 5$^{\rm th}$ line in Tab.~\ref{SLG}). When the relativistic spin-orbit coupling is included, these nodal lines or surfaces (which may be weakly gapped by the spin-orbit coupling) become symmetry-defined Berry-curvature hotspots. This is illustrated in Fig.~\ref{Berry_curv} on relativistic {\em ab initio} band structures of RuO$_2$ and FeSb$_2$ \cite{Smejkal2021b,Smejkal2022,Mazin2021}.

Since $\cal{T}$ is antiunitary in quantum mechanics, the Berry curvature (\ref{Berry-con-cur}) is odd under $\cal{T}$, $\mathcal{T} \bm{\mathcal{B}}_n({\bf k})=-\bm{\mathcal{B}}_n({\bf -k})$. It implies that the integral in Eq.~(\ref{Berry}) vanishes in $\cal{T}$-symmetric band structures. Breaking of $\cal{T}$-symmetry in the band structure of altermagnets is, therefore, the key property that allows for the observation of macroscopic responses, such as the anomalous Hall effect \cite{Smejkal2021b,Smejkal2020}. Recent experiments \cite{Feng2020a} have detected the anomalous Hall effect in RuO$_2$ of a comparable strength to typical Hall signals in ferromagnets. This is consistent with the predicted strong altermagnetic $\cal{T}$-symmetry breaking in the band structure of this compensated collinear magnet \cite{Feng2020a,Smejkal2020} (cf. Tab.~\ref{exp}).

RuO$_2$ is an example in which the lattice of the magnetic Ru atoms alone would have the opposite-spin sublattices connected by a translation. As mentioned above, this symmetry would imply $\cal{T}$-symmetry of the  band structure (and in combination with inversion symmetry of the crystal also spin degeneracy). The presence of the non-magnetic oxygen atoms is, therefore, essential for the $\cal{T}$-symmetry breaking (and spin splitting) in the altermagnetic band structure of RuO$_2$ and, consequently, for the anomalous Hall effect \cite{Smejkal2020}. The term crystal Hall effect \cite{Smejkal2020,Samanta2020,Shao2021a,Mazin2021} was introduced to highlight this feature. One of the implications, unparalleled in the conventional anomalous Hall effect in ferromagnets,  is that the crystal Hall effect in altermagnets is predicted to flip sign not only when reversing the magnetic moments, but also when the symmetry-breaking arrangement of non-magnetic atoms reverses between the two magnetic sublattices \cite{Smejkal2020}. 

Finally, we recall that in 2D systems, Eq.~(\ref{Berry}) turns into a surface integral proportional to the Berry phase, which becomes quantized when the integration covers the full Brillouin zone in 2D insulators \cite{Hasan2010}. The corresponding quantum Hall effect \cite{Klitzing1980} was demonstrated in graphene at room temperature \cite{Novoselov2007} but it requires a strong magnetic field. 
The ferromagnetic quantum anomalous Hall counterpart  \cite{Chang2013} can be observed at zero magnetic field but, so far, has been limited to Kelvin temperatures \cite{Tokura2019,Deng2020}. Since altermagnetism can host the Berry phase phenomena, and can occur in 2D crystals and in insulators, it opens new possibilities in the search for high-temperature  zero-field quantum Hall phenomena. For a further in-depth discussion of Berry phase physics and non-dissipative Hall transport in altermagnets we refer to the recent topical review \cite{Smejkal2021b}.

\section{Research areas}
We now move to the discussion of the potential of altermagnetism in specific areas of condensed-matter research. We start from spintronics in which, besides the anomalous Hall effect, initial theory predictions have been recently supported by  experiments.

\subsection{Spintronics}
 
The $\cal{T}$-symmetry broken electronic structure in ferromagnets is split into majority and minority spin bands. This results in different conductivities of the two spin channels, which makes electrical currents in ferromagnets spin polarized. Passing the spin-polarized current between a reference and a sensing/recording ferromagnetic electrode in a multilayer structure can generate  giant magnetoresistance (GMR), 
tunneling magnetoresitance (TMR), and spin-transfer torque (STT) effects. These principally non-relativistic strong responses underpin reading and writing of information in commercial spintronic devices  \cite{Chappert2007,Ralph2008,Bhatti2017}.

A STT mechanism in Kramers spin-degenerate antiferromagnets,   proposed more than a decade ago  \cite{Nunez2006}, differs fundamentally from STT in ferromagnets. The theoretical model considered a  transmission of a staggered spin polarization from one to the other antiferromagnet, where the spin polarization and the antiferromagnetic orders in the electrodes were all commensurate. This is a subtle, spin-coherent quantum-interference phenomenon relying on perfectly epitaxial commensurate multilayers \cite{Nunez2006,MacDonald2011,Saidaoui2014,Stamenova2017}. Similarly delicate were the proposed GMR/TMR  effects in these antiferromagnetic structures \cite{MacDonald2011}. This may explain why experimentally, the viability of non-relativistic spintronic concepts in the conventional spin-degenerate antiferromagnets has not been demonstrated to date.

As a result, the attention in the research of spintronics in Kramers spin-degenerate antiferromagnets turned to relativistic phenomena \cite{Jungwirth2016}. Anisotropic magnetoresistance, which can be used to detect a 90$^{\circ}$ reorientation of the antiferromagnetic N\'eel vector \cite{Shick2010}, is an example generally applicable to both types of Kramers spin-degenerate antiferromagnets because it is an even-under-$\cal{T}$ macroscopic response (cf. Sec. III.A). A 2$^{\rm nd}$-order magnetoresistance, which can be used to detect a 180$^{\circ}$ reorientation of the N\'eel vector, is an  example of an odd-under-$\cal{T}$ response that occurs only in the second type of Kramers spin-degenerate antiferromagnets with $\cal{T}$-symmetry broken and inversion-symmetry broken band structures \cite{Godinho2018}. These magnetoreristance responses, as well as the spin-orbit torque (SOT) phenomena used to electrically induce the N\'eel vector reorientation \cite{Zelezny2014}, tend to be weak, owing to their relativistic origin.

\onecolumngrid

\begin{figure}[h!]
\begin{center}
\hspace*{-0cm}\epsfig{width=1\columnwidth,angle=0,file=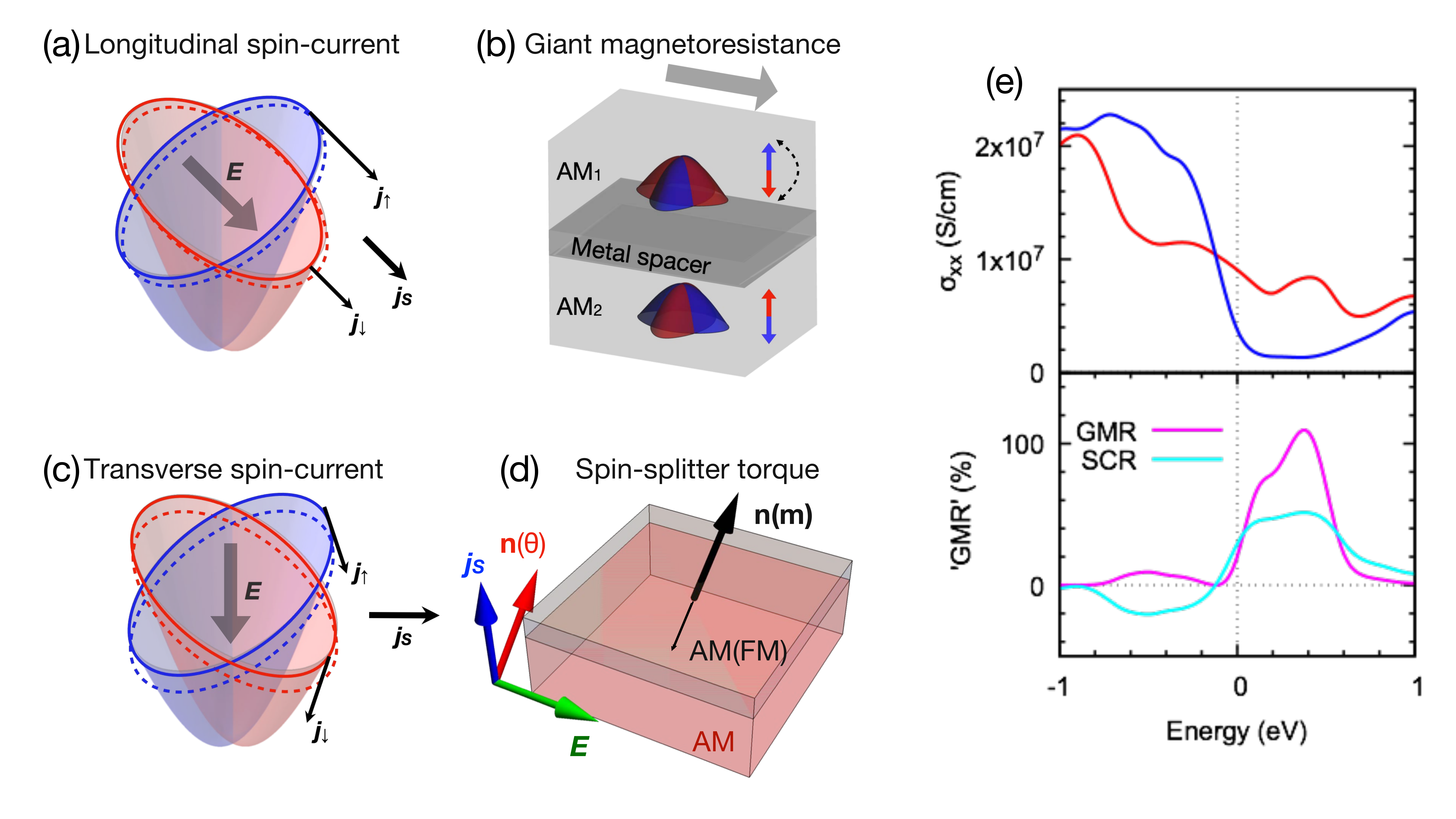}
\end{center}
\vspace{-.5cm}
\caption{
(a) Schematics of the longitudinal spin-current in altermagnets. For an electric bias {\bf E} applied along one of the main anisotropy axes of the spin-split Fermi surfaces, the spin-up and spin-down charge currents are parallel but of different magnitudes due to the Fermi surface anisotropies. As a result, the longitudinal charge current is spin-polarized. (b) Schematics of a GMR stack  in a current-in-plane geometry. As an example, we show the antiparallel configuration of the altermagnetic order vectors in the two  electrodes AM$_1$ and AM$_2$. Interfaces are oriented along one of the main anisotropy axes of the spin-split bands. Energy band cuts highlight the anisotropy around the $\boldsymbol\Gamma$-point, resulting in anisotropic spin-dependent conductivities.  (c) Schematics of the transverse spin-current. For {\bf E} applied in the diagonal direction between the two anisotropy axes of the spin-split Fermi surfaces, the spin-up and spin-down charge currents combine in an unpolarized longitudinal charge current and in a pure transverse spin-current. (d) Spin-splitter-torque concept on  a schematic of altermagnetic  / altermagnetic (ferromagnetic)  bilayer.  A spin-current from the bottom altermagnet propagates in the out-of-plane direction and generates a spin-splitter-torque on the altermagnetic (or ferromagnetic) order vector in top layer. (e) {\em Ab initio} longitudinal spin-up and spin-down conductivities (red and blue), GMR, and the ratio of the transverse spin current relative to the longitudinal charge current (SCR) in RuO$_2$. Adapted from Refs.~\cite{Gonzalez-Hernandez2021,Smejkal2022}.
} 
\label{GMR_fig}
\end{figure}

\twocolumngrid

The strong non-relativistic $\cal{T}$-symmetry breaking and spin splitting in altermagnetic bands directly opens the possibility to not only replicate the concepts known from ferromagnets, but also to enrich non-relativistic spintronics by new effects and functionalities linked to the zero net magnetization \cite{Naka2019,Gonzalez-Hernandez2021,Naka2021,Bose2021,Bai2021,Karube2021,Smejkal2022,Shao2021,Ma2021}.

The anisotropy of the split and equally populated spin-up and spin-down Fermi surfaces in altermagnets (cf. Figs.~\ref{QP}a,b) results in spin-dependent anisotropic group velocities,
$
\partial E_{+}({\bf k})/\partial k_{i} \neq  \partial E_{-}({\bf k})/\partial k_{i},
$
where $+/-$ refers to the spin index.   The corresponding conductivities are also  spin-dependent and anisotropic. Considering $x$ and $y$-direction as the anisotropic axes of the spin-split Fermi surfaces, $\sigma_{+,xx}\neq\sigma_{-,xx}$, $\sigma_{+,yy}\neq\sigma_{-,yy}$, and $\sigma_{\pm,xx}=\sigma_{\mp,yy}$. The electrical current then becomes spin-polarized when the bias is applied along the $x$ or $y$-direction, as schematically illustrated in Fig.~\ref{GMR_fig}a. Moreover, as a consequence of the $\cal{T}$-symmetry breaking of the spin-split bands, the sign of the spin polarization reverses when reversing the altermagnetic order vector. 

The above non-relativistic spin-current characteristics are analogous to ferromagnets. In contrast to ferromagnets, the altermagnetic spin splitting is also predicted to cause the reversal of the  spin polarization of the current when the applied electrical bias is flipped between the $x$ and $y$-direction.

The spin-polarized current directly implies a GMR effect in a stack comprising two altermagnets, separated by a conductive non-magnetic spacer, with the altermagnetic order vectors oriented either parallel or antiparallel, as illustrated in Fig.~\ref{GMR_fig}b. The GMR magnitude can be estimated from the conventional current-in-plane GMR expression derived in ferromagnets \cite{Chappert2007}, 
\begin{equation}
{\rm GMR}=\frac{1}{4}(R_{\sigma}+\frac{1}{R_{\sigma}}-2),
\label{GMR}
\end{equation}
where $R_{\sigma}=\sigma_{+,xx}/\sigma_{-,xx}=\sigma_{+,xx}/\sigma_{+,yy}=\sigma_{-,yy}/\sigma_{-,xx}$. {\em Ab initio} calculations in RuO$_2$
give GMR reaching a 100\% scale \cite{Smejkal2022} (Fig.~\ref{GMR_fig}e), highlighting the expected large GMR ratios in altermagnets.

As noted above, the polarization of the longitudinal spin-polarized current in altermagnets is predicted to reverse not only with the reversal of the altermagnetic order vector but also with the reorientation (e.g. by 90$^\circ$) of the applied electrical bias. A directly related effect, also unparalleled in ferromagnets, is illustrated in 
Fig.~\ref{GMR_fig}c. For a bias applied in the diagonal direction between the two anisotropy axes of the spin-split Fermi surfaces, the longitudinal current is unpolarized. However, a spin-current is generated in the transverse direction.  The effect has been predicted in a range of inorganic and organic materials \cite{Naka2019,Gonzalez-Hernandez2021,Naka2021,Ma2021}.  The altermagnet acts here as an electrical spin-splitter,  with a  propagation angle between spin-up and spin-down currents reaching 34$^\circ$ in RuO$_2$ \cite{Gonzalez-Hernandez2021}. The corresponding charge-spin conversion ratio reaches remarkable 28\% (Fig.~\ref{GMR_fig}e), and the spin-conductivity is a factor of three larger than the record value from a survey of 20,000 non-magnetic relativistic spin-Hall materials \cite{Zhang2021b}.

The outstanding charge-spin conversion characteristics of altermagnetic RuO$_2$ prompt a theoretical proposal of a spin-splitter torque (SST) \cite{Gonzalez-Hernandez2021}, in part already supported by initial experiments \cite{Bose2021,Bai2021,Karube2021} (cf. Tab.~\ref{exp}). In the geometry schematically illustrated in 
Fig.~\ref{GMR_fig}d, an in-plane bias  generates the non-relativistic spin-current in the altermagnetic film along the out-of-plane direction, with the polarization of the spin-current controlled by the orientation of the altermagnetic order vector. The spin current then exerts a torque on the adjacent 
altermagnetic (or ferromagnetic)
layer. SST does not inherit the  problems of STT associated with the out-of-plane direction of the applied electrical bias \cite{Ralph2008,Brataas2012}. Instead, it shares the in-plane electrical-bias geometry of the SOT generated by the relativistic spin-Hall polarizer, while circumventing the limitations of the more subtle relativistic spintronic effects \cite{Manchon2019}.

\begin{figure}[h!]
\begin{center}
\hspace*{-0cm}\epsfig{width=.9\columnwidth,angle=0,file=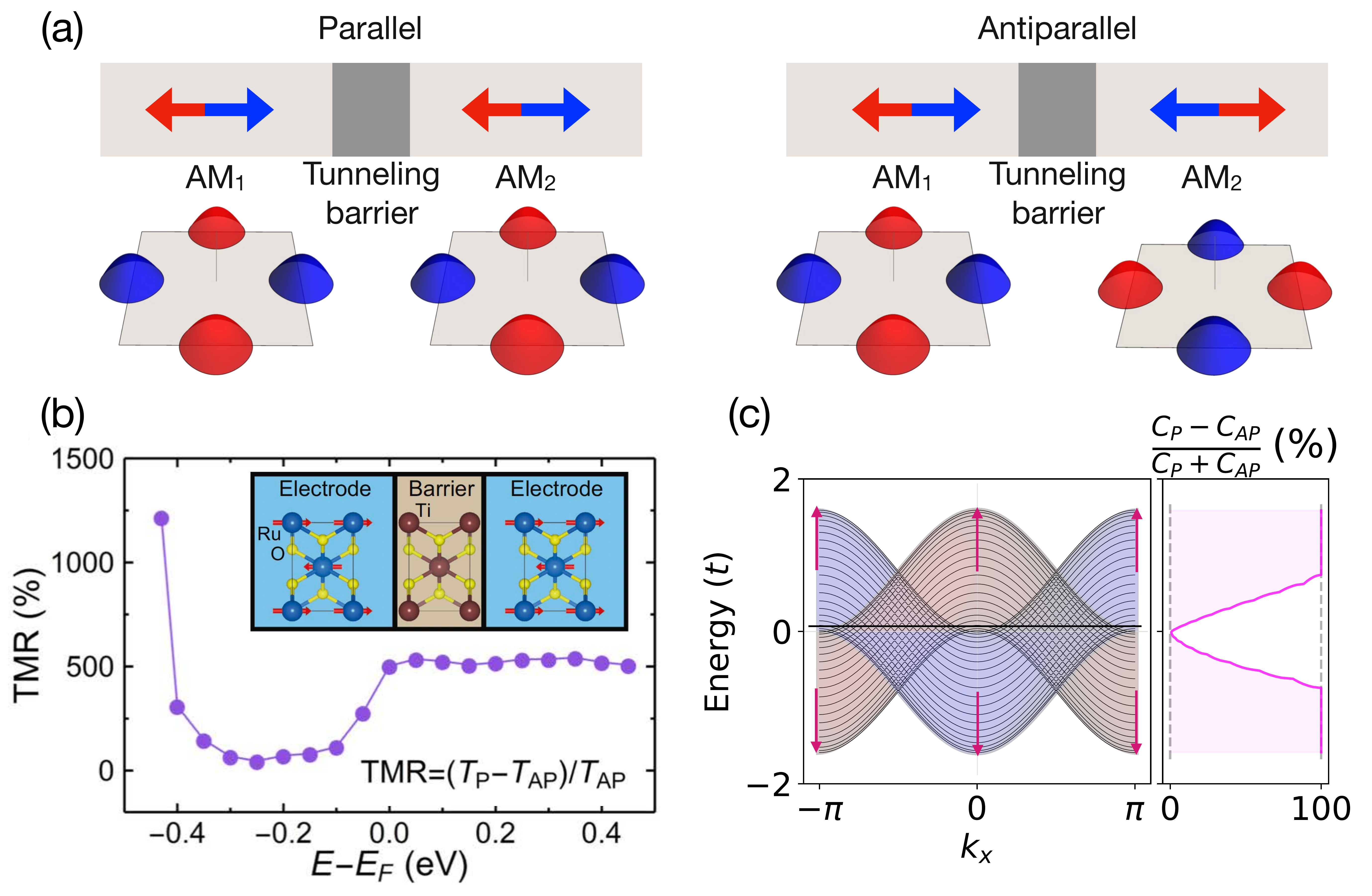}
\end{center}
\vspace{-.5cm}
\caption{
(a) Schematics of a TMR stack with an insulating barrier and altermagnetic electrodes with parallel and antiparallel order vectors.   Energy band cuts highlight the oppositely split valleys, resulting in valley and spin-dependent densities of states. (b) {\em Ab initio} quantum-transmission calculations of the TMR  in a RuO$_2 \vert$ TiO$_2 \vert$ RuO$_2$ tunnel junction. (c) Model spin and transport-momentum projected band structure, and relative difference between the conductances in the parallel and antiparallel configuration of the altermagnetic order vectors in the two electrodes. The TMR is maximized for transport energies corresponding to the spin-split valleys well separated in momentum.  Adapted from Refs.~\cite{Smejkal2022,Shao2021}. 
} 
\label{TMR_fig}
\end{figure}

Another foreseen non-relativistic spintronic effect is an altermagnetic variant of the TMR in a tunnel junction with an insulating spacer separating the two altermagnetic electrodes \cite{Smejkal2022,Shao2021}. The altermagnetic TMR can be illustrated on the model band structure with spin-spit valleys (cf. Figs.~\ref{QP}c,d). The pairs of  valleys with opposite spin polarization result in the equal net population of spin-up and spin-down states, while the densities of states within a given valley become spin dependent, $n_+({\bf M}_1)\neq n_-({\bf M}_1)$,  $n_+({\bf M}_2)\neq n_-({\bf M}_2)$, and $n_{\pm}({\bf M}_1)=n_{\mp}({\bf M}_2)$.  For tunneling which conserves the valley index,  parallel and antiparallel configurations of  altermagnetic order vectors in the two layers, illustrated in Fig.~\ref{TMR_fig}a, are predicted to give different conductances, in analogy to ferromagnetic TMR. This can be seen by applying the Julli\`ere formula \cite{Chappert2007} per valley \cite{Smejkal2022},
\begin{equation}
{\rm TMR}=\frac{1}{2}(R_{n}+\frac{1}{R_{n}}-2),
\label{TMR}
\end{equation}
where the ratio of the spin-up and spin-down densities of states in the valley is given by, $R_{n}=n_+({\bf M}_1)/n_-({\bf M}_1)=n_+({\bf M}_1)/n_+({\bf M}_2)=n_-({\bf M}_2)/n_-({\bf M}_1)$.  

{\em Ab initio} calculations of $\sim 100\%$ TMR ratios in RuO$_2$ (Fig.~\ref{TMR_fig}b) or Mn$_5$Si$_3$ \cite{Smejkal2022,Shao2021} illustrate the potential for achieving large TMR responses in  tunnel junctions with altermagnetic electrodes.

Finally, we note that symmetry-wise, TMR is in principle expected in all altermagnetic spin groups \cite{Smejkal2022}, and can reach large magnitudes as long as the spin-polarized quasiparticles are well separated in the momentum space to provide for the sufficiently decoupled spin transport channels (Fig.~\ref{TMR_fig}c). On the other hand, the  GMR derived from the anisotropy of the macroscopic (averaged over momentum) spin-dependent conductivities, is predicted to be allowed only in selected altermagnetic spin groups \cite{Smejkal2021a,Smejkal2022}. 

\onecolumngrid

\twocolumngrid

\subsection{Ultra-fast optics and neuromorphics}
Achieving energy efficient fast switching is among the key outstanding problems in spintronics \cite{Baumgartner2017,Kimel2019}. The principally isothermal reorientation switching of magnetization in ferromagnets has a threshold when the electrical writing pulse-length is scaled down to the magnetization precession time-scale. This is determined by the inverse of the magnetic resonance frequency which in ferromagnets is typically in the GHz-range. The corresponding threshold writing pulse-length is then in the ns-range.  For longer pulses,  the switching current amplitude is almost independent of the pulse length and, therefore, the writing energy (Joule heating) linearly decreases with decreasing pulse length. Below the threshold, however, the writing current amplitude has to increase to keep the magnetization precession time-scale comparable to the pulse length. The energy in this regime starts to linearly increase with decreasing pulse length, making the switching  prohibitively energy-costly \cite{Bedau2010,Garello2014,Prenat2016,Baumgartner2017}. The $\sim$~ns threshold of electrical writing pulse-lengths in ferromagnets applies to  switching via current-induced Oersted fields, as well as to the spintronic STT or SOT switching mechanisms.

One of the driving ideas behind the antiferromagnetic spintronics research has been the prospect of fast SOT switching, enabled by the  antiferromagnetic resonance frequencies reaching a THz-scale \cite{Jungwirth2016,Olejnik2018}. While in ferromagnets the resonance frequency is determined  by the weak magnetic-anisotropy energy, in antiferromagnets it scales with the geometric mean of the magnetic-anisotropy energy and the strong exchange energy responsible for the antiparallel magnetic order \cite{Kittel1951}. 

Altermagnets can share with antiferromagnets the exchange-enhanced resonance frequency, favorable for achieving energy efficient reorientation switching at pulse lengths well below $\sim$~ns. Unlike antiferromagnets, as discussed in Sec.~IV.A, they are also predicted to enable the favorable non-relativistic  spin-torque mechanisms of  current-induced switching, as well as the readout by the strong non-relativistic GMR/TMR responses.  

A conceptually different approach to achieve higher magnetization dynamic frequencies in ferromagnets, and by this comparably higher writing speeds, is to utilize the large exchange energy by exciting finite wave-vector spin-waves.  The process can be triggered, e.g., by strong heating laser pulses \cite{Beaurepaire1996,Radu2011}, and involves demagnetization of the ferromagnet. This brings us to the frontier research of ultra-fast switching in magnets by optical fs-laser pulses \cite{Kimel2019}. The optical switching mechanisms are principally distinct from the reorientation switching of magnetization by Oersted fields or effective current-induced spin-torque fields. The ultra-fast reorientation switching by fs-laser pulses can involve, e.g., unequal demagnetization/remagnetization dynamics of different spin-sublattices in ferrimagnets, or demagnetization followed by domain-nucleation and expansion in ferromagnets \cite{Kimel2019}.
The fundamental distinction between switching principles when using electrical and optical excitations splits spin-electronics and opto-spintronics in ferromagnets (ferrimagnets) into largely independent research and development fields. 

Recently, materials used in the antiferromagnetic spintronics research have opened a possibility to bridge the electrical and optical switching by a unifying mechanism  that is principally different from all the above mechanisms based on the reorientation of the average magnetic order vector. Over the full range from ms electrical pulses to fs-laser pulses,  the antiferromagnet can be demagnetized and then quenched into a metastable nano-fragmented antiferromagnetic domain state with a higher resistivity than the uniform antiferromagnetic ground state \cite{Kaspar2021}. This quenching of the antiferromagnet into the multi-domain state does not involve a control of the mean orientation of the N\'eel vector. Once the system is quenched into the complex magnetic textures after the electrical or optical excitation, the lack of long-range stray fields in the antiferromagnetic crystals  can inhibit efficient removal of the non-equilibrium magnetic textures  \cite{Kaspar2021}. Since the required switching energy is associated with bringing the system by the electrical or optical excitation to the antiferromagnetic-paramagnetic transition, it does not show the unfavorable increase of the switching energy with decreasing pulse length, typical of the fast current-induced precessional reorientation switching in ferromagnets. 

Resistivity increase in the quenched state on the $\sim10-100\%$ scale, and insensitivity to extreme magnetic fields \cite{Kaspar2021,Zubac2021}, have been associated with the pulse-induced formation of atomically-sharp domain walls in the antiferromagnetic film corresponding to an abrupt flip of the sign of the N\'eel vector between two neighboring atomic planes \cite{Krizek2020a}. Thanks to the latest developments in scanning transmission electron microscopy (STEM) with sub-unit cell spin resolution \cite{Krizek2020a,Kohno2022}, the presence of stationary atomically-sharp domain walls in the antiferromagnet has been visualized by direct imaging \cite{Krizek2020a}. 

While so far explored in the Kramers spin-degenerate antiferromagnets, quenching into the atomic-scale magnetic textures with pulse lengths ranging from ms down to fs-scales may be equally applicable to altermagnets, since they share with antiferromagnets the stray-field-free compensated  magnetic ordering. Moreover, in altermagnets, each atomically-sharp domain wall separating domains with opposite sign of the altermagnetic-order vector can be viewed as a local GMR/TMR junction that is self-assembled, and represents the ultimate down-scaling of the width of the junction spacer. This additional GMR/TMR contributions to the resistances of the atomically-sharp domain walls, absent in the Kramers spin-degenerate antiferromagnets, can be expected to significantly enhance the resistive switching signals of the quenched multi-domain (multi-domain-wall) states in altermagnets.

The above prospect of information coding  down to atomic length-scales and fs time-scale gives a strong incentive for further development of  microscopies with ultra-high spatial and temporal resolution, and with both charge and spin sensitivity. Recently reported STEM images of altermagnetic ferrite Fe$_2$O$_3$ with sub-unit cell spin resolution represent a major step in this direction \cite{Krizek2020a,Kohno2022}. Another recent progress in scanning-probe microscopy imaging of anisotropic charge  distribution on individual atoms \cite{Mallada2021} can facilitate a complementary insight into the microscopic spin-splitting mechanism (cf. 4$^{\rm th}$ line in Tab.~\ref{SLG} and Sec.~III.B), and the corresponding GMR/TMR mechanism in the altermagnet on the nano-scale. High temporal resolution can be achieved in pump-probe fs-laser experiments. They can utilize expected large changes in reflectivity \cite{Kaspar2021} of the altermagnet, that accompany the large quenching-induced resistive changes. Optical reflectivity can substitute here the more elaborate and subtle magneto-optical detection of dynamics of the magnetic-order vector, conventionally employed in  the reorientation  switching experiments \cite{Kimel2019}. 

Apart from the interest in the area of ultra-fast optical switching by fs-laser pulses, the quenching mechanism can have potential applications in neuromorphic computing. The quenching mechanism  facilitates highly reproducible reversible  analog switching and relaxation characteristics, reminiscent of spiking neural-network components \cite{Borders2017,Olejnik2017,Kurenkov2019,Kurenkov2020,Kaspar2021}.
For example, the time dependence in the studied antiferromagnets of the resistivity after the pulse-excitation  has a universal smooth form of a Kohlrausch stretched-exponential relaxation  \cite{Phillips2006}. This can be used in neuromorphic circuits  to mimic the neuron's leaky integration,  or to detect the pulse order and delay that determine the synaptic weights \cite{WulframGerstner2002,Kurenkov2019}. The smooth analog switching and relaxation functions are distinct from ferromagnetic neuromorphic  devices whose characteristic behavior are stochastic fluctuations between two states with opposite magnetization \cite{Grollier2016a,Kaiser2022}.

Finally, we point out that the foreseen altermagnetic neuromorphic devices with the ms-fs scalability of the writing pulse-length, reversible and reproducible analog time-dependent switching and relaxation characteristics, and with strongly enhanced resistive switching signals by the local GMR/TMR at atomically-sharp domain walls, represent an attractive complementary research direction to the more traditional charge-based neuromorphic devices. Charge memristors led to successful demonstrations of analog synapses in proof-of-concept artificial neural networks \cite{Li2018}. However, the ionic transport nature of their operation, that facilitates large resistive switching signals and the memristive characteristics, also imposes principle endurance and speed limitations \cite{Yang2013}. These limitations are absent in the spintronic neuromorphic devices, including the foreseen altermagnetic neuromorphic components, as they rely on the manipulation of the charge and spin of electrons.

\subsection{Thermoelectrics, field-effect electronics and multiferroics}

The key merit of ferromagnets from energy-saving perspective is non-volatility, i.e., that they can store information even when power is switched off. On the other hand, electrical reading and especially writing information into ferromagnetic memory devices can generate significant Joule heating  \cite{Chappert2007}. This can be directly harvested during the writing process  in which the elevated temperature effectively reduces the equilibrium energy barrier separating the states with opposite magnetization orientation.  In the latest generation of hard disks, elevating the temperature of a bit while recording is provided through an external laser heat-source. 
Similarly, the all-optical switching by laser pulses via the demagnetization processes mentioned in Sec.~IV.B is typically accompanied by significant heating effects. This brings us to a discussion of how altermagnetism, rather the generating heat, can contribute to energy harvesting in devices combining heat, charge and spin phenomena. 

Ferromagnets have been considered for a direct heat conversion to electricity \cite{Bauer2012b}. Here the anomalous Nernst effect, a thermo-electric counterpart of the anomalous Hall effect, is regarded as an attractive candidate phenomenon \cite{Mizuguchi2019}. The anomalous Nernst effect generates an electric field in a transverse direction to the thermal gradient. Particularly in thin-film or nanostructured heat-charge conversion device, the transverse geometry can significantly enhance the conversion efficiency compared to the conventional longitudinal Seebeck effect \cite{Mizuguchi2019}.  A complementary research area to the anomalous Nernst effect are thermal counterparts of the GMR/TMR and STT phenomena. Here the energy harvesting concept is based on employing heat gradients, instead of electrical bias, to directly read or write information in a memory device \cite{Bauer2012b}.

Altermagnets significantly enlarge the material landscape for realizing and optimizing these thermo-electric responses that originate, in analogy to their electronic counterparts, from the $\cal{T}$-symmetry broken spin-split band structure. Unlike the typically metallic ferromagnets, altermagnets are predicted to span a broad range of conduction types (cf. Sec.~II.D). This is favorable because, from a general thermo-electric perspective, semimetals or semiconductors are more suitable material types than metals due to the strong dependence of their electronic structure on energy near the Fermi level. A particularly intriguing case in this context are materials showing a metal-insulator transition. Among the altermagnetic candidate materials, FeSb$_2$ \cite{Smejkal2021a,Mazin2021} is an example in which earlier studies reported an extraordinary (spin-independent) thermo-electric response, linked to the metal-insulator transition \cite{Jie2012}. 

Similar to thermoelectrics, the non-metallic materials are favorable for field-effect electronics. A research of ferromagnetic semiconductors, in particular of the Mn-doped III-V compounds, was motivated by the prospect of integrating spintronics with field-effect transistor functionalities in one material \cite{Ohno1998}. Because magnetism is carrier-mediated in these materials, one of the driving ideas in this area of research was the control of the magnetic order by electrostatic gating effects \cite{Dietl2014}. Research of devices based on the III-V semiconductors also led to discoveries of the spin Hall effect and SOT, that gave birth to the field of relativistic spintronics \cite{Sinova2015,Manchon2019}. These successes, on one hand, and the failure to achieve high ferromagnetic transition temperatures in these semiconductors, on the other hand, provided one of the key incentives driving the research field of antiferromagnetic relativistic spintronics \cite{Jungwirth2016}. Altermagnets open a new prospect of materials combining not only high magnetic transition temperatures with non-metallic band structures, but also with the strong non-relativistic spintronic responses generated by the $\cal{T}$-symmetry breaking and spin splitting in the energy bands.

Multiferroic materials \cite{Ramesh2007a} can complement the magnetic semiconductors by offering a non-volatile electric control of magnetism via the internal coupling between the ferroelectric and magnetic orders. Only insulating materials can be ferroelectric, otherwise the free carriers would screen out the electric polarization. This again disfavors ferromagnets that are mostly metallic. The prominent multiferroic materials are non-centrosymmetric perovskite oxides with a compensated magnetic order \cite{Ramesh2007a}. As shown in Sec.~II.D, altermagnetism is compatible with this material family. Here CaMnO$_3$ is an example multiferroic  \cite{Bhattacharjee2009}, that is also a material candidate for hosting the altermagnetic phase \cite{Smejkal2021a}. 

\subsection{Superconductivity}
The family of insulating perovskite oxides brings us to its prominent cuprate member La$_2$CuO$_4$ that, upon doping, turns into  a high-temperature $d$-wave superconductor  \cite{Si2016}. The recognition that this cuprate crystal belongs to an altermagnetic spin group with a $d$-wave character of the spin-momentum locking  \cite{Smejkal2021a} brings us to foreseen research of the interplay of altermagnetism and superconductivity \cite{Smejkal2021a,Mazin2022}. Research in this context may include areas such as the coexistence of altermagnetism  and unconventional superconductivity with anisotropic Cooper pairing \cite{Sigrist1991}, altermagnetic fluctuations as a pairing mechanism \cite{Sigrist1991}, or phenomena at altermagnet/superconductor interfaces \cite{Mazin2001,Flensberg2021}. 

Since altermagnets have spin-degenerate nodal lines or surfaces protected by the spin-group symmetries, a spin-singlet Cooper pairing may occur for the corresponding momenta, in analogy to conventional antiferromagnets. For the spin-singlet case, the $2\times2$ Cooper-pair potential matrix (gap function or order parameter), $\hat{\boldsymbol\Delta}({\bf k})$, satisfies $\Delta_{\uparrow\uparrow}({\bf k})=\Delta_{\downarrow\downarrow}({\bf k})=0$,  $\Delta_{\downarrow\uparrow}({\bf k})=-\Delta_{\uparrow\downarrow}({\bf k})$, and $\Delta_{\uparrow\downarrow}({\bf k})=\Delta_{\uparrow\downarrow}(-{\bf k})$ \cite{Sigrist1991}. The matrix is unitary, with corresponding zero net spin average of the pairing state, and describes even-parity wave Copper pairing, including the anisotropic, e.g., $d$-wave pairing.

On the other hand, the altermagnetic spin-group symmetries also allow for spin-split and broken $\cal{T}$-symmetry parts of the Brillouin zone where, $\epsilon(s,{\bf k})\neq\epsilon(-s,-{\bf k})$ and $A{\bf H}\,{\bf k}\neq{\bf k}$ (cf. 3$^{\rm rd}$ and 6$^{\rm th}$ line in Tab.~\ref{SLG}). These momenta can support spin-triplet Cooper pairing. In analogy to ferromagnets, the spin-triplet Cooper-pair potential matrix corresponding to a spin-split spin-up Fermi surface of  the altermagnet, $\hat{\boldsymbol\Delta}^{(\uparrow)}({\bf k})$, takes a form, $\Delta^{(\uparrow)}_{\uparrow\downarrow}({\bf k})=\Delta^{(\uparrow)}_{\downarrow\uparrow}({\bf k})=\Delta^{(\uparrow)}_{\downarrow\downarrow}({\bf k})=0$, and $\Delta^{(\uparrow)}_{\uparrow\uparrow}({\bf k})=-\Delta^{(\uparrow)}_{\uparrow\uparrow}(-{\bf k})$ \cite{Sigrist1991}. The matrix in this case is non-unitary and describes odd-parity wave Copper pairing \cite{Sigrist1991}. Unlike ferromagnets, however, the  altermagnetic spin-group symmetries impose the presence of a counter-part spin-down Fermi surface with a corresponding  $\hat{\boldsymbol\Delta}^{(\downarrow)}({\bf k^{\prime}})$ that satisfies, $\Delta^{(\downarrow)}_{\uparrow\downarrow}({\bf k^{\prime}})=\Delta^{(\downarrow)}_{\downarrow\uparrow}({\bf k^{\prime}})=\Delta^{(\downarrow)}_{\uparrow\uparrow}({\bf k^{\prime}})=0$, $\Delta^{(\downarrow)}_{\downarrow\downarrow}({\bf k^{\prime}})=-\Delta^{(\downarrow)}_{\downarrow\downarrow}(-{\bf k^{\prime}})$, and $\Delta^{(\downarrow)}_{\downarrow\downarrow}({\bf k^{\prime}})=\Delta^{(\uparrow)}_{\uparrow\uparrow}({\bf k})$, where ${\bf k^{\prime}}=A{\bf H}\,{\bf k}$. On one hand, altermagnets can thus share with ferromagnets  the spin-triplet symmetry of Cooper pairing while, unlike ferromagnets on the other hand, a zero net spin average of the spin-triplet superconducting state is protected by the altermagnetic spin-group symmetries. In the context of unconventional supercoductivity, the  ferromagnetic-antiferromagnetic dichotomy of altermagnetism, as well as the features unparalleled in neither the conventional ferromagnets or antiferromagnets,  mirror the discussion in the previous sections. 

Apart from the compatibility of altermagnetism with the different types of Cooper pairing, altermagnetic fluctuations can provided earlier unexplored mechanisms for generating the pairing. Since the electron-phonon coupling mechanism tends to be limited to the conventional spin-singlet $s$-wave pairing \cite{Sigrist1991}, altermagnets can be particularly relevant for research of the unconventional superconductivity, including both the spin-singlet and spin-triplet anisotropic types of Cooper pairing. 

Finally, we foresee intriguing new physics at altermagnet/superconductor interfaces in areas including Andreev reflection \cite{Mazin2001} or Majorana fermion quasiparticles \cite{Flensberg2021}.  On one hand, the behavior of alternagnets at these interfaces can be reminiscent of conventional antiferromagnets when dominated by the spin-symmetry protected nodal lines or surface. On the other hand, interface orientations exposing the strong altermagnetic spin splitting can generate a phenomenology similar to  the ferromagnet/superconductor interfaces. As in the case of bulk crystals, the research of interface effects can exploit the predicted broad range of altermagnetic material types. 

\section{Conclusion}
\begin{figure}[h!]
\begin{center}
\hspace*{-0cm}\epsfig{width=.8\columnwidth,angle=0,file=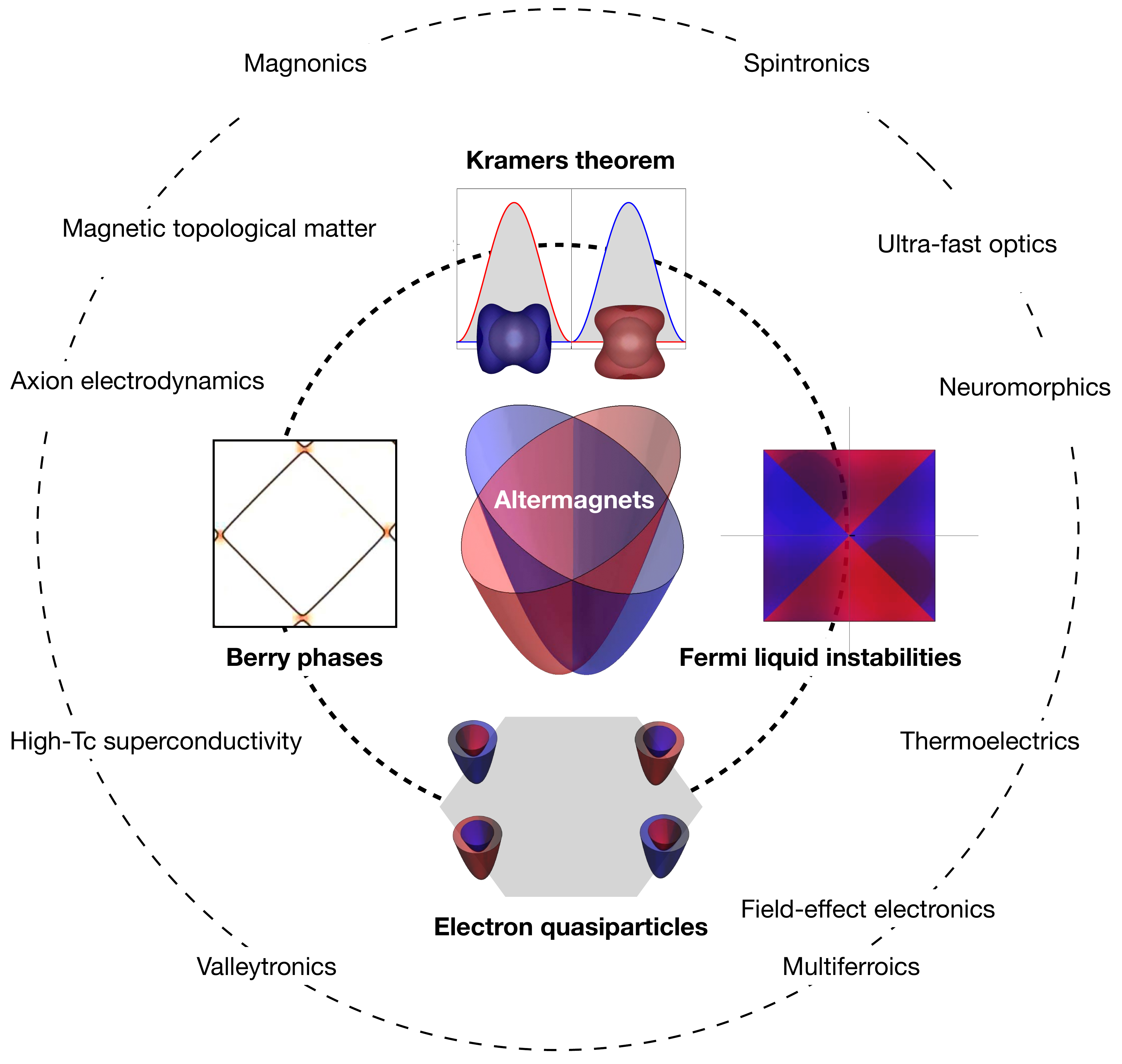}
\end{center}
\vspace{0cm}
\caption{Summary of the emerging research landscape of altermagnetism. 
} 
\label{summary}
\end{figure}

In this Perspective we have drawn a picture of a third basic magnetic phase that emerges on the fundamental level of non-relativistic uncorrelated band-theory  of non-frustrated collinear magnets. The altermagnetic phase can be uniquely defined in both  crystal-structure real space and  electronic-structure momentum space,  systematically classified and described by symmetry-group theory, and is predicted to be abundant among diverse material types. Most importantly, the notion and the significance of a distinct phase becomes apparent from the unique ways in which altermagnetism can contribute to the development of fundamental physical concepts, and to the research in modern condensed-matter physics fields. Given the still relatively early stage of our understanding of altermagnetism, and the limited space, our choice of the discussion topics  in this Perspective should be regarded as broadly illustrative and provisional. We can anticipate that in the near future, altermagnetism will have impact in other fields  including magnetic topological matter, magnonics, valleytronics, or axion electrodynamics. Fig.~\ref{summary} summarizes the emerging research landscape of altermagnetism.

\subsection*{Acknowledgement}
We acknowledge fruitful interactions with Igor Mazin. This work was supported by  Ministry of Education of the Czech Republic Grants LNSM-LNSpin, LM2018140,
Czech Science Foundation Grant No. 19-28375X,  EU FET Open RIA Grant No. 766566, SPIN+X (DFG SFB TRR 173) and Elasto-Q-Mat (DFG SFB TRR 288). We acknowledge the computing time granted on the supercomputer Mogon at Johannes Gutenberg University Mainz (hpc.uni-mainz.de).

\onecolumngrid

\newpage

\twocolumngrid

\end{document}